\documentclass[a4paper, 10pt]{elsarticle}

\usepackage{amsmath}
\usepackage{amssymb}
\usepackage{graphicx}
\usepackage{subcaption}
\usepackage{url}
\usepackage[top=3em, bottom=5em, left=3em, right=3em]{geometry}
\usepackage{tabularx}
\usepackage{supertabular}
\usepackage{multirow}
\usepackage{boxedminipage}
\usepackage{eurosym}
\usepackage{xcolor}
\usepackage{soul}
\usepackage{mathtools}
\usepackage{bm}
\usepackage{flushend}

\newtheorem{proposition}{Proposition}

\begin{document}

\newcommand{\uBinary}{\ensuremath{u_{m}}}
\newcommand{\bBinaryTau}{\ensuremath{b^\tau_{m,i}}}
\newcommand{\bBinaryFlowMax}{\ensuremath{b^F_{m,j}}}

\newcommand{\priceDemand}{\ensuremath{p^D_{t,k}}}
\newcommand{\priceDemandTilde}{\ensuremath{\widetilde{p}^D_{t,k}}}
\newcommand{\priceGen}{\ensuremath{p^G_{t,p}}}
\newcommand{\priceGenTilde}{\ensuremath{\widetilde{p}^G_{t,p}}}

\newcommand{\setDemandNode}{\ensuremath{\Omega^D_{t,n}}}
\newcommand{\setDemandAll}{\ensuremath{\Omega^D_{t}}}
\newcommand{\setGenNode}{\ensuremath{\Omega^G_{t,n}}}
\newcommand{\setGenAll}{\ensuremath{\Omega^G_{t}}}

\newcommand{\setNodes}{\ensuremath{\mathcal{N}}}
\newcommand{\setLines}{\ensuremath{\mathcal{M}}}
\newcommand{\setLoops}{\ensuremath{\mathcal{L}}}
\newcommand{\setTime}{\ensuremath{\mathcal{T}}}
\newcommand{\setTariff}{\ensuremath{\mathcal{I}}}
\newcommand{\setLumpyInvestments}{\ensuremath{\mathcal{J}}}
\newcommand{\setLumpyCapExpansionsF}{\ensuremath{\mathcal{\bar{F}}_m}}

\newcommand{\discountFactor}{\ensuremath{\omega_t}}

\newcommand{\PTDF}{\ensuremath{\delta_{m,n}}}

\newcommand{\demand}{\ensuremath{d_{t,k}}}
\newcommand{\gen}{\ensuremath{g_{t,p}}}

\newcommand{\demandMax}{\ensuremath{d^{max}_{t,k}}}
\newcommand{\genMax}{\ensuremath{g^{max}_{t,p}}}

\newcommand{\bFunc}{\ensuremath{B_k(\demand)}}
\newcommand{\cFunc}{\ensuremath{C_p(\gen)}}

\newcommand{\flow}{\ensuremath{f_{t,m}}}
\newcommand{\flowMax}{\ensuremath{F_{m}}}
\newcommand{\flowMaxDiscrete}{\ensuremath{\bar{F}_{m,j}}}
\newcommand{\flowMaxCap}{\ensuremath{\hat{F}_{}}}
\newcommand{\flowZero}{\ensuremath{F^0_m}}

\newcommand{\KFix}{\ensuremath{K^{fix}_{m}}}
\newcommand{\KVar}{\ensuremath{K^{var}_{m}}}

\newcommand{\tariff}{\ensuremath{\tau_m}}
\newcommand{\tariffDiscrete}{\ensuremath{\bar{\tau}_{m,i}}}

\newcommand{\incidenceMatrixElement}{\ensuremath{a_{m,n}}}
\newcommand{\sensitivityMatrixElementKVL}{\ensuremath{\psi_{\ell,m}}}

\newcommand{\price}{\ensuremath{\pi_{t,n}}}
\newcommand{\muMax}{\ensuremath{\mu^{max}_{t,m}}}
\newcommand{\muMin}{\ensuremath{\mu^{min}_{t,m}}}
\newcommand{\gammaKVL}{\ensuremath{\gamma_{t,\ell}}}
\newcommand{\phiDemMax}{\ensuremath{\varphi^{D,max}_{t,k}}}
\newcommand{\phiGenMax}{\ensuremath{\varphi^{G,max}_{t,p}}}

\newcommand{\auxBTauDemand}{\ensuremath{y^{b^\tau d}_{t,m,i,k}}}
\newcommand{\auxBTauGen}{\ensuremath{y^{b^\tau g}_{t,m,i,p}}}
\newcommand{\auxBFMuMaxU}{\ensuremath{y^{b^F\mu,max}_{t,m,j}}}
\newcommand{\auxBFMuMinU}{\ensuremath{y^{b^F\mu,min}_{t,m,j}}}

\newcommand{\Tau}{\mathrm{T}}

\newtheorem{remark}{Remark}

\begin{frontmatter}

	\title{Ex-ante dynamic network tariffs for transmission cost recovery \tnoteref{ack}}

	\author[bath]{Iacopo~Savelli\corref{cor1}}
	\ead{is572@bath.ac.uk}
	
	\author[bath]{Antonio~De~Paola}
	\ead{adp50@bath.ac.uk}
	
	\author[bath]{Furong~Li}
	\ead{eesfl@bath.ac.uk}

	\cortext[cor1]{Corresponding author}
	\tnotetext[ack]{This work was supported by the EPSRC under Grant EP/S00078X/1 and by the
	Leverhulme Trust under Grant ECF-2016-394}
	
	\address[bath]{Department of Electronic and Electrical Engineering, 
		University of Bath, UK}
	
	\begin{abstract}
		This paper proposes a novel tariff scheme and a new optimization framework in order to address the recovery of fixed investment costs in transmission network planning, particularly against rising demand elasticity. At the moment, ex-post network tariffs are utilized in addition to congestion revenues to fully recover network costs, which often leads to over/under fixed cost recovery, thus increasing the investment risk. Furthermore, in the case of agents with elastic market curves, ex-post tariffs can cause several inefficiencies, such as mistrustful bidding to exploit ex-post schemes, imperfect information in applied costs and cleared quantities, and negative surplus for marginal generators and consumers. These problems are exacerbated by the increasing price-elasticity of demand, caused for example by the diffusion of demand response technologies. To address these issues, we design a dynamic ex-ante tariff scheme that explicitly accounts for the effect of tariffs in the long-term network planning problem and in the underlying market clearing process. Using linearization techniques and a novel reformulation of the congestion rent, the long-term network planning problem is reformulated as a single mixed-integer linear problem which returns the combined optimal values of network expansion and associated tariffs, while accounting for price-elastic agents and lumpy investments. The advantages of the proposed approach in terms of cost recovery, market equilibrium and increased social welfare are discussed qualitatively and are validated in numerical case studies.	
	\end{abstract}
	
	
\end{frontmatter}

\twocolumn
	
\section*{Nomenclature}

\subsection*{\textbf{Sets and Indices}}

\begin{supertabular}{l p{0.8\columnwidth}}
	$\setLumpyCapExpansionsF$ & set of possible lumpy capacity expansions for line $m$, with $\setLumpyCapExpansionsF = \bigcup_{j \in \setLumpyInvestments} \flowMaxDiscrete$\\
	$i$ & index of a discrete tariff level, with $i \in \setTariff$\\
	$\setTariff$ & set of discrete tariff levels\\
	$j$ & index of a lumpy capacity expansion, with $j \in \setLumpyInvestments$\\
	$\setLumpyInvestments$ & set of lumpy capacity indices\\
	$k$ & index of a consumer, with $k \in \setDemandAll$\\
	$\ell$ & index of a loop, with $\ell \in \setLoops$\\
	$\setLoops$ & set of loops\\
	$m$ & index of a line, with $m \in \setLines$\\
	$\setLines$ & set of lines\\
	$n$ & index of a node, with $n \in \setNodes$\\
	$\setNodes$ & set of nodes\\
	$p$ & index of a producer, with $p \in \setGenAll$\\
	$t$ & index of a time period, with $t \in \setTime$\\
	$T$ & number of time periods\\
	$\setTime$ & set of time periods, with $\setTime$ = \{1, ..., T\}\\
	$\setDemandNode$ & set of consumers in node $n$\\
	$\setDemandAll$ & set of all consumers, i.e., $\setDemandAll = \cup_n \setDemandNode$\\
	$\setGenNode$ & set of producers in node $n$ \\
	$\setGenAll$ & set of all producers, i.e., $\setGenAll = \cup_n \setGenNode$\\
\end{supertabular}

\subsection*{\textbf{Parameters}}
\begin{supertabular}{l p{0.8\columnwidth}}
$\incidenceMatrixElement$ & element of the network incidence matrix $A_{\setLines\times\setNodes}$. It holds $a_{m,n}=1$ ($a_{m,n}=-1$) if the positive power flow on line $m$ exits (enters) from node $n$. In all other cases, $a_{m,n}=0$ \\
$\demandMax$ & maximum quantity demanded by consumer $k$ at time $t$\\
$\flowZero$ & existing capacity on the line $m$\\
$\flowMaxDiscrete$ & lumpy capacity expansion for line $m$, with $j \in \setLumpyInvestments$\\
$\genMax$ & maximum quantity offered by producer $p$ at time $t$\\
$\KFix$ & fixed cost of building or expanding line $m$\\
$\KVar$ & variable cost of building or expanding line $m$\\
$\priceDemandTilde$ & bid demand price by consumer $k$ at time $t$ without network costs\\
$\priceGenTilde$ & bid offer price by producer $p$ at time $t$ without network costs\\
$\PTDF$ & allocation factor for cost distribution among market participants\\
$\tariffDiscrete$ & discretized transmission tariff applied on line $m$, with $i \in \setTariff$\\
$\sensitivityMatrixElementKVL$ & element of the sensitivity matrix $\Psi_{\setLoops \times \setLines}$ representing the reactance of line $m$ in loop $\ell$.\\
$\discountFactor$ & weighting factor.\\
\end{supertabular}

\subsection*{\textbf{Functions}}
\begin{supertabular}{l p{0.8\columnwidth}}
$\bFunc$ & benefit function for consumer $k$\\
$\cFunc$ & cost function for producer $p$.\\
\end{supertabular}

\subsection*{\textbf{Variables}}
\begin{supertabular}{l p{0.8\columnwidth}}
	$\bBinaryFlowMax $ & binary variable equal to one if the lumpy investment in additional capacity $\flowMaxDiscrete$ for line $m$ is made, and zero otherwise\\
	$\bBinaryTau$ & binary variable equal to one if the tariff $\tariffDiscrete$ is applied on line $m$, and zero otherwise\\ 
	$\demand$ & accepted demand for consumer $k$ at time $t$\\
	$\flow$ & flow in the line $m$ at time $t$\\
	$\gen$ & accepted generation for producer $p$ at time $t$\\
	$\priceDemand$ & bid demand price by consumer $k$ at time $t$ including network costs\\
	$\priceGen$ & bid offer price by producer $p$ at time $t$ including network costs\\
	$\tariff$ & transmission tariff applied on line $m$ \\
	$\uBinary$ & binary variable equal to one if line $m$ is expanded, and zero otherwise.\\
\end{supertabular}

\subsection*{\textbf{Dual variables and associated constraint}}

\begin{supertabular}{l p{0.8\columnwidth}}
	$\price$ & power balance constraint at time $t$ in node $n$\\
	$\gammaKVL$ & Kirchhoff's voltage law constraint for loop $\ell$ at time $t$\\
	$\muMax$ & maximum flow constraint at time $t$ in line $m$\\
	$\muMin$ & minimum flow constraint at time $t$ in line $m$\\
	$\phiDemMax$ & maximum demanded quantity constraint for consumer $k$ at time $t$\\ 
	$\phiGenMax$ & maximum offered quantity constraint for producer $p$ at time $t$\\ 
\end{supertabular}

\subsection*{\textbf{Auxiliary variables}}

\begin{supertabular}{l p{0.8\columnwidth}}
	$\auxBFMuMaxU$ & it replaces the product $\bBinaryFlowMax\muMax$\\
	$\auxBFMuMinU$ & it replaces the product $\bBinaryFlowMax\muMin$\\
	$\auxBTauDemand$ & it replaces the product $\bBinaryTau\demand$\\
	$\auxBTauGen$ & it replaces the product $\bBinaryTau\gen$\\
\end{supertabular}

\section{Introduction}

In the near future, an unprecedented amount of network investments will be necessary to accommodate the substantial shifts in generation and demand caused by increasing penetration of renewable energy resources,  and electrification of transportation and heating \cite{europeanparliament2017investments}. The current costs due to congestions in the electrical network in Europe amount to more than \euro2.4 billion and are expected to further increase. For this reason, a significant number of grid expansion projects are planned in the short and medium term \cite{entsoe2018BiddingZoneConf}. Globally,  the estimated cost of the planned transmission investments in the next 15 years amounts to \$ 1.7 trillion \cite{energyAgency2014worldOutlook}.
In this scenario, it is of paramount importance that current paradigms for transmission network planning are properly updated to sustain the development of a sound, secure, and affordable energy system. 

A comprehensive survey of the main challenges in transmission network planning, from both an economic and an engineering point of view, is reported in \cite{wu2006Transmission}. 
\textcolor{black}{Some of the most relevant issues that are currently being tackled include:  i) combining the planning of transmission expansions with new synchronous plants \cite{Barati2015} and additional wind generation \cite{Gu2012}, ii) accounting for planning uncertainties \cite{Moreira2018} and iii) developing new investment paradigms for transmission expansion in deregulated power markets. An overview of this last point is provided in \cite{transmissionPlanningUnderCompetitiveElectricityMarketDavid2001} and  some market and game-theory approaches are presented in \cite{fonseka07OptTransmissionExpansion} and \cite{dePaolaMerchantGame2018}, respectively.}

A wide array of different planning methodologies is also being considered. Reference \cite{farrell2018auction} proposes a novel framework to incorporate transmission planning decisions in pay-as-bid auctions (such as those for competitive renewable energy procurement) in order to obtain a reduction of the long-term total system cost. 
Reference \cite{gan2019security} presents a security-constrained co-planning of transmission line expansion and energy storage in the presence of high penetration of wind power. The proposed approach is based on a Bender decomposition algorithm, where the master problem handles the planning phase, whereas the sub-problem introduces corrective post-contingency actions to minimize the costs under fault conditions. 
Reference \cite{pozo2017doing} proposes a proactive method in planning for transmission network expansion by taking into account also generation expansion decisions. The resulting formulation is cast as a mathematical program with equilibrium constraints, which is solved with a specifically developed algorithm. 
An interesting analysis of how different network configurations and transmission expansion planning decisions can result in different attributions of carbon emission responsibilities to consumers is reported in reference \cite{sun2017analysis}.
Uncertainty in network planning is another important issue that may lead to inefficient decisions, in particular in the context of very high renewable energy penetration. In this sense, reference \cite{sun2018novel} proposes a data-driven scenario generation framework for transmission expansion to account for unseen but important load and wind power scenarios while considering inter-spatial dependencies between loads and wind generation outputs, by using a vine-copula approach. 
Note that, the presence of tariffs and price-sensitive market players can have an impact not only on transmission planning but also on generation expansion planning, which is a strictly connected problem \cite{koltsaklis2018state}, as well as on ancillary service markets \cite{koltsaklis2018incorporating}.

\subsection{Motivation}

A key general aspect of transmission expansion planning that still exhibits several unresolved issues is the full and efficient recovery of the investment costs and in particular of fixed costs \cite{kishore2015analysis}. This is a well-known problem that has been analysed extensively \cite{borenstein2018DoTE}. One of the first proposed solution from a general economic perspective has been the introduction of Ramsey pricing \cite{ramsey1927contribution}, which has been further analysed in a power system context by Boiteux in \cite{boiteux1960peak} and \cite{boiteux1971management}. However, this approach only considers monopolies in a fully-regulated environment and does not include market interactions between generation and demand. \textcolor{black}{Several alternative approaches have been proposed, such as two-part tariffs  \cite{littlechild1975TwoPartTariffs} and non-linear pricing methods \cite{laffont1998PriceDiscrimination}. Despite these contributions, the recovery of fixed costs remains a topic with several open questions, such as the impact of new technologies on network charging \cite{pollitt2018NetworkChargingPPS} and a fair allocation of the network costs among the different network participants \cite{hogan2018Primer}.}

Currently, network expansion costs are recovered through congestion rent and tariffs. At the moment, network charges are not harmonized between countries, and usually each nation applies its own tariff scheme, which can significantly differ from the others. In this sense, a comprehensive review of the different tariff approaches adopted in Europe and North America is presented in \cite{ACERreport} and \cite{LUSZTIG2006}. The specific examples of TNUoS in UK and TUoS in Australia are discussed in \cite{NG_TNUoS} and \cite{AEMO_TUoS}, respectively. These tariffs are generally levied on a capacity or energy basis and each country has set different allocations of charges between generation and load. A subset of countries, such as Great Britain, Ireland and Norway, consider locational signals and envisage different payments for generator and demand, depending on the area in which they operate.

A fundamental point is that all the discussed network tariff schemes are applied ex-post, without accounting for their effect on both price-sensitive consumers and producers \cite{CUSC2019}. This means that the clearing process in electricity markets does not properly consider the effect of network tariffs on the accepted quantities and on the market clearing prices. This approach can lead to several inefficiencies, such as cleared orders which result in a negative surplus and mistrustful biddings due to the risk of distorting the participants behaviour \cite{bushnell2014EfficiencyTrasmissionCost}. 
These issues are becoming particularly relevant nowadays, because demand is no longer perfectly inelastic and consumers are moving towards a flexible and price-sensitive behaviour. \textcolor{black}{This has motivated further analyses on the impact of demand elasticity on pricing \cite{Chicco2000} and has led to the design of new tariffs \cite{schittekatte2018TariffDesign} and network charges \cite{abdelmotteleb2018NetworkCharges} for systems with active, price-responsive customers}. The points highlighted above become even more relevant if one considers that the necessity of new network investments is substantially increasing the tariffs for the consumers. In the case of the UK, it is expected that the overall tariff payments will increase by $\pounds 800$ m over the next five years, with a $25\%$ to $50\%$ increase of the tariffs for commercial and domestic users \cite{NG_TNUoS_5y}.

\subsection{Novel Contributions}
Given the substantial increase in both demand flexibility and tariffs payments in the near future, it is crucial to establish more efficient methods for allocation and recovery of transmission costs. In this context, the paper presents a novel framework for optimal network expansion that utilizes ex-ante tariffs to ensure full recovery of investment costs (both their fixed and variable components). To the best of the authors' knowledge, this represents the first comprehensive effort in transmission network tariff design to account for price-sensitive market participants by considering the tariff components directly into the investment optimization and market clearing process.
This allows to properly accommodate bids and offers of price-elastic agents (i.e. generators and demand) which under the current ex-post tariffs may be subject to sub-optimal market outcomes and allocation inefficiencies \cite{bushnell2014EfficiencyTrasmissionCost}. These results are obtained by characterizing the network expansion problem as a non-linear mixed-integer bilevel optimization. In this formulation, the upper level depicts the long-term investment planning problem: the overall social welfare is maximized while recovering investment costs from tariffs (determined dynamically) and congestion rent. In turn, the lower level  represents a short-term market clearing problem which takes into account the presence of network tariffs. By using standard integer algebra, complementarity properties, and a novel reformulation of the congestion rent, the proposed bilevel model is recast as a mixed-integer linear program (MILP) which can be solved with off-the-shelf solvers without resorting to any iterative algorithm. 

To summarize, the main novelties of the proposed approach are:
\begin{itemize}
\item utilization of ex-ante network tariffs, to properly account for price-elastic market participants;
\item dynamic optimal distribution of investment cost recovery between congestion rent and tariffs, with improvements on overall social welfare;
\item new formulation of the network planning problem as a bilevel optimization and design of equivalent MILP formulation to facilitate numerical resolution;
\item a novel characterization of the congestion rent, which can be recast as a linear expression of dual variables when lumpy investments are considered.
\end{itemize}

\subsection{Paper structure}
The remaining part of this paper is organized as follows. Section~\ref{sec:issues} describes the main issues in transmission investment cost recovery. In particular, the following problems are discussed: i) the recovery of fixed costs, ii) the presence of lumpy network investments, and iii) the effect of tariffs on price-sensitive market participants. Section \ref{sec:ex-anteTariffs} and \ref{sec:novelParadigm} introduce the proposed formulation of ex-ante tariffs and the bilevel model for network planning, respectively. Section~\ref{sec:resolutionMethod} recasts this model as an equivalent, single MILP problem and Section~\ref{sec:numericalResults} presents two case studies to show the main properties of the proposed approach. Finally, Section~\ref{sec:conclusion} outlines the main conclusions.

\section{Transmission Planning: model and challenges}
\label{sec:issues}
The problem of network expansion is commonly analysed under a regulated centralized paradigm \cite{fonseka07OptTransmissionExpansion}. The network expansion is determined by a central network planner with the purpose of maximizing the total social welfare of the system. This scheme can be formulated analytically through the following optimization problem  \cite{dePaolaMerchantGame2018}:
\begin{subequations}\label{CS}
	\begin{align}
	\max_{\substack{\demand, \gen,\\ \uBinary, \flow, \flowMax}} & \sum_{t \in \setTime} \Big(\sum_{k \in \setDemandAll} \bFunc - \sum_{p \in \setGenAll} \cFunc\Big) \notag\\
	&- \sum_{m \in \setLines} \uBinary\Big(\KFix + \KVar\flowMax\Big) \label{CSLevelObjectiveFunction} \\
	\text{s.t.}&\notag\\
	&\sum_{k \in \setDemandNode}\demand - \sum_{p \in \setGenNode}\gen + \sum_{m \in \setLines} \incidenceMatrixElement \flow = 0  \notag\\
	& \hspace{11em} \forall t \in \setTime, \forall n \in \setNodes \label{CSPowerBalance}\\
	&\sum_{m \in \setLines} \sensitivityMatrixElementKVL \flow = 0 \hspace{3.1em} \forall t \in \setTime, \forall \ell \in \setLoops \label{CSKVLContraint}\\
	&\flow \leq \flowZero + \uBinary\flowMax \hspace{2.6em} \forall t \in \setTime, \forall m \in \setLines \label{CSflowMaxConstraint}\\
	&-\uBinary\flowMax -\flowZero \leq \flow \hspace{1.5em} \forall t \in \setTime, \forall m \in \setLines \label{CSflowMinConstraint} \, ,
	\end{align}
\end{subequations}
with $\uBinary \in \{0,1\}$, $0 \leq \demand \leq \demandMax$ , $0 \leq \gen \leq \genMax$, $\flow \in \mathbb{R}$, and $\flowMax \geq 0$. 
With this approach, the central planner must determine which lines $m$ are expanded (through the binary variable $\uBinary$) and the amount of additional capacity on these lines (through the continuous variable $\flowMax$). The values of $\uBinary$ and $\flowMax$ are selected in order to maximize the objective function in (\ref{CSLevelObjectiveFunction}), which represents the total social welfare of the system and corresponds to the difference between two main components: 
i) short-term benefits, equal to the difference between demand benefits $B$ and generation costs $G$, and ii) long-term investment costs, i.e. the costs of expanding each line $m$ by an additional capacity $\flowMax$. The investment costs include a fixed component 
$\KFix$ (incurred if $\uBinary=1$ and line $m$ is expanded) and a variable component $\KVar \flowMax$ (proportional to the additional capacity $\flowMax$). 
The social welfare optimization is subject to the power balance constraint (\ref{CSPowerBalance}) at each node $n$  and the Kirchhoff's voltage law (\ref{CSKVLContraint}) on each loop $l$ of the considered network. In addition, constraints \eqref{CSflowMaxConstraint} and \eqref{CSflowMinConstraint} set line flow limits, imposing that the power flow $\flow$ (at each time period $t$ and on each line $m$) does not exceed the maximum line capacity $\flowZero+\flowMax$ available after the network expansion, where $\flowZero$ is the existing capacity.
	
This centralized formulation, which is commonly considered a starting point in analytical studies on network planning, exhibits some relevant drawbacks, which are discussed in the rest of this section. 

\subsection{Recovery of fixed and variable investment costs}
\label{sec:fixedcosts}
The network expansion investment can be recovered through the congestion rent (CR) collected on the transmission lines. Denoted by $\price$ the locational marginal price at node $n$ at time $t$, i.e. the Lagrange multiplier associated to constraint (\ref{CSPowerBalance}), the CR can be expressed as follows:
\begin{equation}
\label{eq:CR}
CR = \sum_{t \in \setTime} (\pi_{t,n^r_m}-\pi_{t,n^s_m}) \flow
\end{equation}
where $n^s_m$, $n^r_m \in \setNodes$ denote the sending and receiving node of branch $m$, respectively. A key issue of recovering costs exclusively through the congestion rent CR is that such approach does not recover fixed costs, even tough the objective function \eqref{CSLevelObjectiveFunction} accounts for both the fixed and the variable investment costs. The reason is that the solution obtained in \eqref{CS} maximizes the social welfare of all the participants, which is composed by three components: i) the consumer surplus, ii) the producer surplus, and iii) the congestion rent \cite{Chicco2000}.  
If the optimal solution of \eqref{CS} envisages a certain network expansion (i.e. $\uBinary=1$), it follows that the resulting investment cost is lower than the associated social welfare increase. However, there is no guarantee that the cost (in both its fixed and variable components) will be entirely covered by the CR, which only constitutes one component of the social welfare.
For the specific case with $\flowZero=0$ (to neglect previous investments), the total congestion rent CR in (\ref{eq:CR})
is exactly equal to the variable costs, regardless of the network topology \cite{kirschen2004fundamentals}: 
\begin{equation}
CR = \sum_{m \in \setLines} \KVar\flowMax.
\end{equation}
As a consequence, when the transmission system operator is entitled to collect only the congestion rent, the fixed costs $\KFix$ of the optimal planned network expansion returned by \eqref{CS} cannot be recovered, despite the fact that the expansion actually generates sufficient welfare to cover all the costs.
  
\begin{figure}[h!]
\centering
\includegraphics[width=0.4\textwidth]{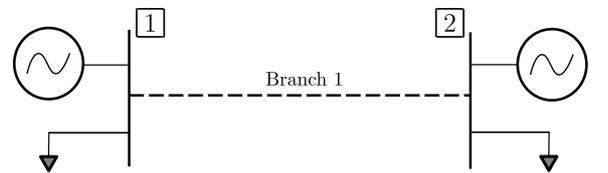}
\caption{Topology of the 2-node system.}
\label{fig:2node}
\end{figure}

\begin{figure}[h!]
	\centering
	\begin{subfigure}{.45\textwidth}
		\centering
		\includegraphics[width=1\linewidth]{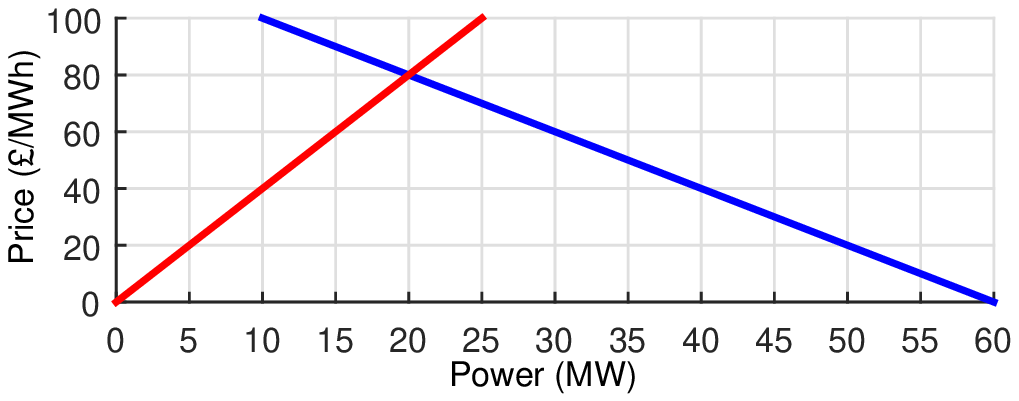}
		\label{fig:zone1}
	\end{subfigure}
	\begin{subfigure}{.45\textwidth}
		\centering
		\includegraphics[width=1\linewidth]{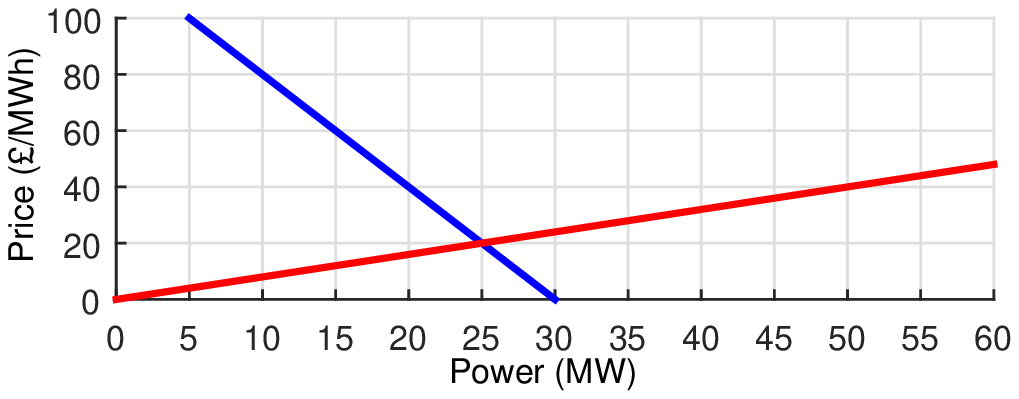}
		\label{fig:zone2}
	\end{subfigure}
	\caption{Demand curve in blue and supply curve in red for first (left) and second (right) market zones.}
	\label{fig:zone12}
\end{figure}

To further emphasize this point, consider problem (\ref{CS}) over a single time period for the simple 2-node network in Fig.~\ref{fig:2node}, whose demand and supply curves are depicted in Fig.~\ref{fig:zone12}.

\begin{figure}[h!]
	\centering
	\includegraphics[width=0.45\textwidth]{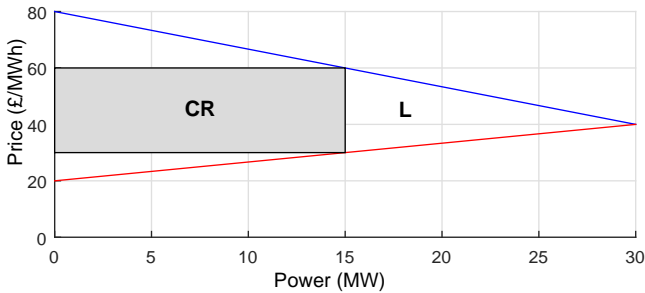}
	\caption{The blue curve represents the import demand curve, given by the horizontal difference between the demand curve and supply curve in zone 1. The red curve depicts the export supply curve, given by the horizontal difference between the supply curve and demand curve in zone 2. The grey area is the congestion rent, whereas the triangular area marked by L is the dead-weight loss due to the flow limits.}
	\label{fig:IDES}
\end{figure}

\begin{figure}[h!]
	\centering
	\includegraphics[width=0.45\textwidth]{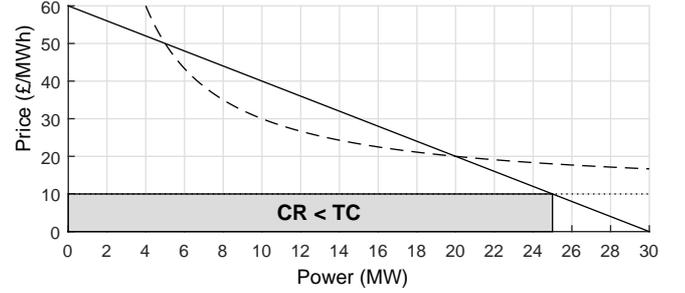}
	\caption{The solid line represents the price difference between the import and export curves in Fig.~\ref{fig:IDES}. The horizontal dotted line is the average variable costs (i.e. $\KVar=10$ \pounds/MWh), whereas the dashed hyperbole depicts the ATC in (\ref{eq:ATC}).}
	\label{fig:netIDES}
\end{figure}

If no connection exists between the two zones, the market clearing price in the first zone is $80$~\pounds/MWh, and $20$~\pounds/MWh in the second zone. In the first zone, for each price lower than $80$~\pounds/MWh, some consumers would be willing to buy, but not enough producers are willing to produce at these prices. However, this excess demand can be satisfied by importing energy from the second zone. Define the import curve for the first zone as the horizontal (energy) difference between the demand and supply curves in the first zone. This curve is depicted as the blue line in Fig.~\ref{fig:IDES} for prices lower than $80$~\pounds/MWh. Similarly, in the second zone, for each price greater than $20$~\pounds/MWh, some producers would be willing to produce, but not enough consumers are willing to buy. This excess supply can be exported. Define the export curve as the horizontal (energy) difference between the supply and the demand curve in the second zone. This amount is depicted as a red line in Fig.~\ref{fig:IDES} for prices greater than $20$~\pounds/MWh. The curves reported in Fig.~\ref{fig:IDES} are useful to represent the congestion rent. 
Suppose now that the two zones are connected by a transmission line with a maximum capacity of $15$ MW. Assuming a time period of 1 hour, the energy traded between the two zones is $15$ MWh. In this case, the congestion rent is represented by the grey area in Fig.~\ref{fig:IDES}, where the market price is equal to $60$~\pounds/MWh in the first zone and $30$~\pounds/MWh in the second zone. The triangular part above the congestion rent is the welfare increase due to the importing consumers, whereas the triangular part below the congestion rent represents the increase in welfare due to the exporting producers. The triangular area labelled by L is the dead-weight loss, i.e., the loss of social welfare caused by the flow limit.
The solid line in Fig.~\ref{fig:netIDES} represents the difference between the price curves depicted in Fig.~\ref{fig:IDES}. It means that the same deductions in terms of congestion rent and dead-weight loss can be obtained by using this curve. 

\begin{figure}[h!]
	\centering
	\includegraphics[width=0.45\textwidth]{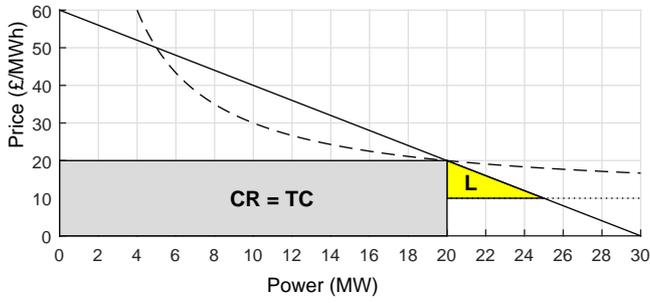}
	\caption{This figure depicts the same curves as in Fig.~\ref{fig:netIDES}. The grey rectangular area and the triangular yellow area correspond to the congestion rent CR and the value loss L.
	}
	\label{fig:netIDES2}
\end{figure}

For this simple case with a single time period, 
the condition $\flow=\flowMax$ must hold \cite{kirschen2004fundamentals}, and relationship between congestion rent and investment costs can be graphically represented as in Fig.~\ref{fig:netIDES}.
In model \eqref{CS}, the total investment cost (TC) corresponds to the last term in (\ref{CSLevelObjectiveFunction}) and is defined as $TC=\sum_{m \in \setLines} \uBinary(\KFix + \KVar\flowMax)$. In this two-zone example, the average total cost (ATC) per unit of added line capacity (if the investment is made, i.e. $\uBinary=1$), is defined as follows:
\begin{equation}\label{eq:ATC}
ATC = \dfrac{TC}{\flowMax} = \dfrac{\KFix}{\flowMax} + \KVar
\end{equation}
For $\KFix=200$ \pounds/MWh and $\KVar=10$ \pounds/MWh, the ATC are depicted as the dashed hyperbole in  Fig.~\ref{fig:netIDES}, whereas the horizontal dotted line represents the average variable costs, that are constant and equal to $\KVar$ in problem \eqref{CS}. 
The optimal capacity expansion for the optimization problem \eqref{CS} corresponds to the x-value of the intersection between the average variable costs (dotted line) and the price differential among the two zones (solid line) when the added line capacity is equal to $\flowMax$. In the considered case, the optimal line expansion is equal to $25$ MW and the associated price difference between the two zones is $10$~\pounds/MWh.
The congestion rent and the variable costs are both represented by the grey rectangular area, and take the same value of $250$~\pounds.
However, at this optimal solution of (\ref{CS}), the ATC  (dashed hyperbole in Fig. \ref{fig:netIDES}) is equal to $18$~\pounds/MWh and the total costs amount to $450$~\pounds. As previously discussed, the investment in this case does not recover the fixed costs $\KFix$, equal to $200$~\pounds. In order to recover both the fixed and variable costs, the following \textit{revenue adequacy} condition must be added to problem \eqref{CS}:
\begin{equation}\label{revenueAdequacyIntro}
CR \geq TC = \sum_{m \in \setLines} \uBinary(\KFix + \KVar\flowMax) \, .
\end{equation} 
The solution of (\ref{CS}) with this additional constraint is represented in Fig.~\ref{fig:netIDES2}. In this case, the optimal expansion corresponds to the x-value of the intersection between the price differential curve (solid line) and the ATC (dashed line), which also accounts for averaged fixed costs. The optimal value implies an energy exchange of $20$ MWh. 
The rectangular grey area is the congestion rent, which is exactly equal to the total costs, and amounts to $400$~\pounds.  The consequence of constraint \eqref{revenueAdequacyIntro} is a reduction of the inter-zone exchange from $25$ to $20$ MWh. The triangular yellow area, labelled by L, is the reduction in the objective function. 
The revenue adequacy constraint, by imposing full recovery of the investment costs through congestion rent, leads to underinvestment (from $\flowMax=25$ MW to $\flowMax=20$ MW), and to a reduction of the overall social welfare of $25$~\pounds.

\subsection{Lumpy investments}
\label{sec:lumpy}
\begin{figure}[h!]
	\centering
	\includegraphics[width=0.45\textwidth]{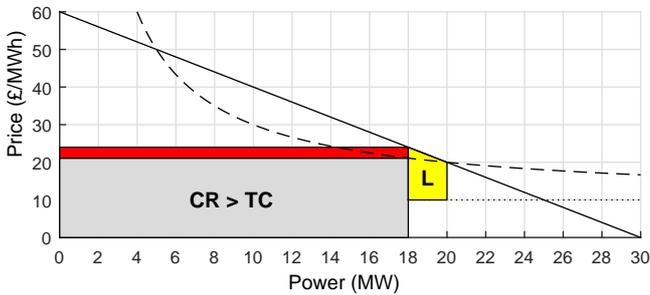}
	\caption{Effect of lumpy investment. The traded energy is limited to $18$ MWh. The total cost are represented by the are labelled by TC. The transmission system operator collects an extra-profit from the congestion rent, equal to the red area on top of the total costs. The decrease in the objective function is represented by the yellow area labelled by L.}
	\label{fig:Lumpy}
\end{figure}

In practice, the capacity expansion on transmission lines cannot take any arbitrary value and, instead, can only correspond to a finite set of possible discrete options. This complicates the formulation and resolution of problem (\ref{CS}), as the continuous decision variables $\flowMax$ are replaced by binary variables, with a significant computation burden increase. Moreover, lumpiness of capacity expansions exacerbates the issue of fixed cost recovery and sub-optimality of the resulting solution. Consider again the example presented in Section \ref{sec:fixedcosts} and assume now that the feasible line expansions must take values in the finite set $D=\{ 3~\text{MW}, 6~\text{MW}, \dots, 30~\text{MW} \}$, with a discrete step of $3$ MW between each option. This means that the previous optimal solution of $20$ MW is not feasible any more. In this case of lumpy investments, the optimal solution corresponds to further under-investment and a capacity expansion of $18$ MW, as depicted in Fig.~\ref{fig:Lumpy}.
The TC are represented by the grey area. Note that the transmission system operator collects an extra-profit from the congestion rent, equal to the red area above the total costs. The further decrease in the objective function is represented by the yellow area L, and amounts to $24$~\pounds.  

\begin{proposition}
In the presence of fixed costs, if the transmission operator revenue is limited to congestion rent, the optimal planning problem can lead to a sub-optimal state for consumers and producers. Lumpy investment exacerbate this issue. However, consumers and producers could be willing to bear some costs to compensate the network operator, as long as the achievable solution leads to a greater surplus.
\end{proposition}

\subsection{Network tariffs and price-sensitive market participants}
\label{sec:expost_tariffs}
In the previous section, it has been shown that the recovery of fixed costs in transmission planning problems through congestion rent leads to under-investments and suboptimal solutions. To avoid these undesired outcomes, one alternative option is to directly recover these costs from generators and consumers, which have benefited from the social welfare increase of the investment. This is generally obtained by introducing ex-post tariffs, i.e. additional payments from generation and demand after the market clearing. However, this solution also presents relevant drawbacks and it can lead to suboptimal solutions. Consider in fact that demand and supply market curves in day-ahead markets are price-elastic \cite{demandResponseSuKirshen2009}. That is, consumers and producers buy and sell different quantities at different prices. For consumers, the greater the price, the smaller the quantity they are willing to buy. Conversely, for producers, the greater the price, the greater the quantity they are willing to supply. A fundamental consequence of having market participants with price-sensitive bids is that any additional cost (applied after the market clearing process has been determined) can lead to a negative surplus for some participants. 
\begin{figure}[h!]
	\centering
	\includegraphics[width=0.45\textwidth]{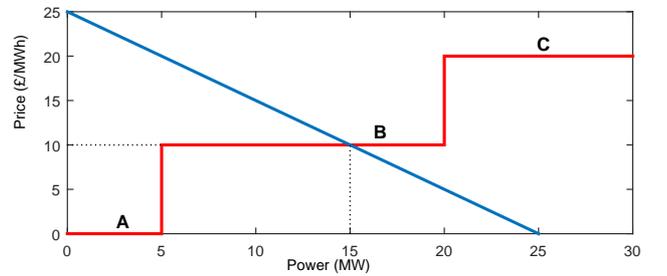}
	\caption{Single-zone market with demand depicted as a blue line, and supply represented as a stepwise red curve. The supply curve is determined by the bids submitted to the market by three producers, labelled by A, B, and C, respectively.  The intersection of the demand and supply curves yields the market price, i.e., $10$~\pounds/MWh. Assuming perfect competition, the producers submitted bids are their marginal costs. Therefore, the profit of producer B is zero, as it both collects and pays $10$~\pounds/MWh.}
	\label{fig:Elastic}
\end{figure}

To see this, consider the single-zone market depicted in Fig.~\ref{fig:Elastic}. The stepwise red curve represents the supply market curve, and it is composed by the three bids submitted by the producers A, B, and C. Assuming perfect competition, the bids represent the (constant) marginal costs of the producers, and are equal to 0, 10, and 20 \pounds/MWh for A, B and C, respectively. The blue line depicts the demand curve.  The market price is determined by the intersection of the demand and supply curves, and is equal to $10$~\pounds/MWh. All the offered quantities by producer A are accepted and it sells $10$ MWh of energy. Given the market price of $10$~\pounds/MWh and the marginal cost of zero, the profit of producer A is $100$~\pounds. By contrast, the quantity offered by producer C is not accepted, because its marginal cost is strictly greater than the market price, formally, the bid order of producer C is out-of-the-money \cite{iacopoAPEN2018}. Producer B is the marginal producer \cite{kirschen2004fundamentals}, and its bid order price (i.e., its marginal cost) sets the market price. This means that producer B collects $100$~\pounds~ as revenues for selling $10$ MWh at $10$~\pounds/MWh, but it pays $10$~\pounds/MWh for each unit produced, therefore its profit is zero. This holds for any marginal producer with constant marginal cost. Now, suppose that an ex-post network tariff $\Tau = 5$~\pounds~ is charged to producers. As a consequence of this extra cost, producer B pays $105$~\pounds~ and collects only $100$~\pounds. In other words, the ex-post network tariff translates directly into a $5$~\pounds~ loss.  
It could be argued that, if producer B knew in advance about the tariff $\Tau=5$~\pounds, it could rise the submitted price to $10.5$~\pounds/MWh for its traded energy of $10$ MWh. However, this implies a priori knowledge of the market clearing solution, which is not the case in a properly functioning market with price-taker participants. 
These results hold for any marginal entity, and can be extended to a more general settings. Recall in fact that, under a nodal pricing framework, if there are $n$ congested lines, then there will be $n+1$ marginal units \cite{litvinov2010locationalmarginalprices}. More generally, if ex-post tariffs are applied, any units whose surplus is smaller than the ex-post tariff will incur in a loss. 

\begin{proposition}\label{prop:2}
	As long as market participants have elastic market curves, network tariffs should be considered directly into the clearing process. This would allow to properly account for consumers and producers price-sensitive behaviour, and it would avoid the execution of orders which could results (ex-post) in a negative surplus.
\end{proposition}

\section{Ex-ante dynamic network tariffs}\label{sec:ex-anteTariffs}

To overcome the drawbacks of ex-post tariffs, discussed in Section \ref{sec:expost_tariffs}, a new type of \textit{ex-ante tariff} is presented. This novel cost-recovery scheme is directly considered into the market clearing process and is related to the traded quantities.

Under the proposed framework, the function $\Tau_p(\gen)$ represents the network costs charged to generator $p$ and $\cFunc$ corresponds to its generation cost function. Then, the total costs to be paid by $p$ are $\cFunc + \Tau_p(\gen)$ and the marginal total  cost $\priceGen$ is defined as follows:
\begin{equation}\label{newMarginalCost}
\priceGen(\gen) =  \dfrac{\partial \cFunc}{\partial \gen} + \dfrac{\partial\Tau_p(\gen)}{\partial \gen}
\end{equation}
where $\gen$ is the produced quantity. The marginal total cost $\priceGen$ represents the bid price the producer $p$ should submit to the market under perfect competition. The key feature of this novel formulation is that, by considering $\Tau_p$, the effect of network charges is directly embodied in the producer's bid price.

Similarly, assume that the function $\Tau_k(\demand)$ represents the network costs charged to consumer $k$, and $\bFunc$ its benefit function before network charges. Then, the net benefit for the consumer $k$ amounts to $\bFunc - \Tau_k(\demand)$. In this case, the actual marginal benefit $\priceDemand$ is:
\begin{equation}\label{newMarginalBenefit}
\priceDemand(\demand) =  \dfrac{\partial \bFunc}{\partial \demand} - \dfrac{\partial\Tau_k(\demand)}{\partial \demand} 
\end{equation}
where $\demand$ is the cleared demand. The value $\priceDemand$ represents the bid price the consumer $k$ should submit to the market under perfect competition.

The fundamental aspect of this new approach is that the effect of network costs is considered \textit{before} the actual market clearing takes place, through the inclusion of ex-ante tariffs $T_p$ and $T_k$. Note that these tariffs directly affect the submitted bid prices, which in turn can modify the market equilibrium (as long as market participants are price-sensitive). This means that tariffs must be determined by using an optimization planning problem that considers also the market clearing process, to properly account for their effect on both prices and quantities. 

\begin{figure}[h!]
	\centering
	\includegraphics[width=0.45\textwidth]{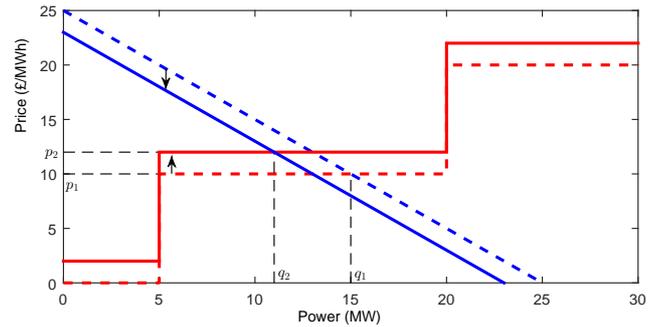}
	\caption{Assume generation marginal costs as depicted by the red dashed line, marginal benefit before network costs as represented by the dashed blue line, and a network tariff of $2$~\pounds. Then, the resulting demand and supply curves are depicted by the blue and red solid lines, respectively.}
	\label{fig:curvesShifted}
\end{figure}

In the proposed formulation (fully detailed in Section~\ref{sec:UpperLevel}), the charged network costs are expressed as linear functions of generation and demand, with $\Tau_p(\gen) \coloneqq \sum_{m \in \setLines} \tariff \gen$, and   $\Tau_k(\demand) \coloneqq \sum_{m \in \setLines} \tariff \demand$, where $\tariff$ is the tariff applied to line $m$. Therefore, the conditions \eqref{newMarginalCost} and \eqref{newMarginalBenefit} can be recast as follows:
	\begin{equation}
	\priceGen =  \priceGenTilde + 	\sum_{m \in \setLines} \tariff \, , \qquad \qquad	\priceDemand =  \priceDemandTilde - \sum_{m \in \setLines} \tariff
	\end{equation}
That is, the bid profile of the generator $p$ is obtained by shifting its marginal costs over the price axis by a positive offset, whereas the bid profile of the consumer $k$ is obtained by shifting its gross marginal benefit over the price axis by a negative offset to account for tariffs to be paid, as depicted in Fig.~\ref{fig:curvesShifted}.
The initial market clearing solution ($q_1,p_1$) with market price $p_1= 10$~\pounds/MWh and accepted power $q_1=15$~MW, is changed by the introduction of tariffs, and takes the new values ($q_2,p_2$) for quantity and power, respectively, as depicted in Fig.~\ref{fig:curvesShifted}. Note how the introduction of tariffs reduces the cleared quantity. This quantity belongs to price-sensitive consumers and producers who suffer a loss when the tariffs are applied ex-post, and therefore it is not accepted when the tariffs are considered ex-ante.
	
The key point of the proposed scheme is to provide a mathematical approach to exactly embed network costs (e.g. tariffs) into participants' bid profiles, which are then reflected into the cleared quantities and market prices. In turn, quantities and market prices are used to compute the congestion rent collected by the transmission system operator, which is therefore affected by the selected tariffs. This means that, in the presence of price-sensitive market participants, accepted quantities, market prices, tariffs and congestion rent are dynamically linked, and all these values must be considered simultaneously when a network investment planning problem is solved. For this reason, the proposed scheme is termed \textit{ex-ante dynamic network tariffs}.

\begin{proposition}\label{propTariffAsVariableCosts}
	Network costs, such as tariffs, should result in marginal costs of the form as in \eqref{newMarginalCost} and \eqref{newMarginalBenefit}, in order to be directly representable by their submitted market bid orders before the actual market clearing takes place.
\end{proposition}

\section{Network planning problem}\label{sec:novelParadigm}
In order to address the issues of traditional network planning approaches, highlighted in Section \ref{sec:issues}, a novel planning paradigm is presented. The purpose of this new formulation is to determine network expansions that maximize the social welfare of the system while explicitly accounting for price-elastic participants and lumpiness of investments. Ex-ante dynamic network tariffs are used to ensure optimal recovery of both fixed and variable investment costs. 

\subsection{Bilevel program for network planning}\label{sec:BilevelProgramming}
In order to account for ex-ante dynamic tariffs, the proposed network planning model is structured as a bilevel program \cite{bard1998practical}. A bilevel model can be regarded as two nested optimization problems, termed upper and lower level problem. Formally, it is defined as follows:
\begin{subequations}\label{sketchBilevel}
	\begin{align}
	\max_{u \in \, \mathcal{U}} &\,F(u,x^*)\label{sketchUpper}\\
	\text{s.t.}\,\,\,&x^* \in \arg \max_{x \in \mathcal{X}} f(x;u) \, ,\label{sketchLower}
	\end{align}
\end{subequations}
where $F$ is the objective function of the upper level problem \eqref{sketchUpper}, $f$ is the objective function of the lower level problem \eqref{sketchLower}, and $\mathcal{U}$, $\mathcal{X}$ are constraint sets. The variables $x^*$ represent the optimal solution of the lower level problem, which depends on the upper level variables $u$, i.e., $x^*\,{=x^*(u)}$. One important characteristic of a bilevel model is that all the upper level variables $u$ enter the lower level as parameters. This feature can be exploited to recast a bilevel model as a single mathematical optimization program when specific conditions hold \cite{conejo2012complementarity}, as detailed in Section \ref{sec:SingleLevel}. Bilevel programming is extensively used in the field of game theory to model Stackelberg problems \cite{stackelberg1934}. However, in power system economics, nested optimization structures are often used to access dual variables \cite{blanco2014consumer,Feijoo2014, iacopoAPEN2018, iacopoCommunityAPEN2019}, which are related to market prices according to the marginal pricing theory \cite{schweppe1988spot}.

\begin{figure}[h!]
	\centering
	\includegraphics[width=0.45\textwidth]{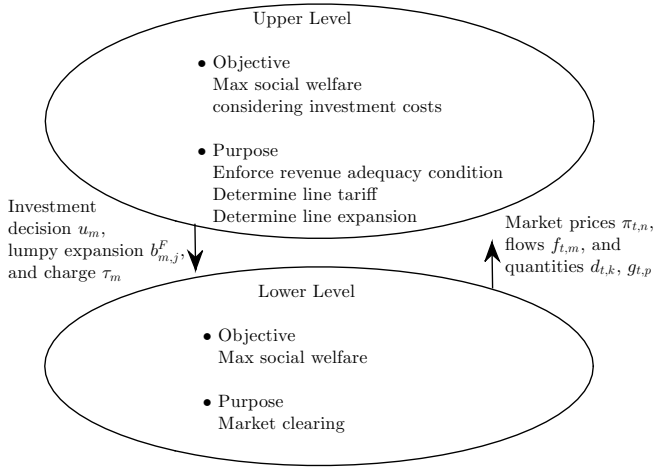}
	\caption{Bilevel structure of the proposed formulation for optimal network planning. The upper level program represents the long term investment planning problem. Given the optimal decision variables of the upper level, which include the investment decision $\uBinary$, the selected lumpy expansion $\bBinaryFlowMax$, and the tariff $\tariff$, the lower level problem actually performs a market clearing problem. As a result of the clearing process the prices $\price$, the flows $\flow$, and the executed quantities $\demand$ and  $\gen$ are obtained, which are used by the upper level to verify the revenue adequacy constraint.
	} 
	\label{fig:Bilevel}
\end{figure}

The proposed bilevel framework for optimal network planning, with dynamic ex-ante tariffs for full cost recovery, is sketched in Fig.~\ref{fig:Bilevel}. The lower level represents the day-ahead market clearing problem, whereas the upper level represents the long term investment planning problem.
The objective of the upper level is to determine the optimal investment plan by considering both the social welfare and the investment costs, while enforcing the revenue adequacy condition, i.e. the recovery of both fixed and variable costs. In the proposed framework, the transmission operator collects both the congestion rent, and the tariffs applied to consumers and producers. The key decision variables of the upper level problem are: i) the investment decision $\uBinary$, ii) the selected lumpy capacity expansion $\bBinaryFlowMax$, and iii) the tariffs $\tariff$. Having fixed these quantities, the lower level problem actually clears the market, while directly accounting for network tariffs into the clearing process. As a result of the clearing process, the market prices $\price$, the flows $\flow$, the executed quantities $\demand$ and  $\gen$ are obtained, which are used by the upper level problem to compute the transmission operator revenues, which are compared with the investment costs to verify the revenue adequacy condition.

\vskip 0.2cm
The fundamental feature of the proposed approach is the explicit inclusion of the network tariffs $\tariff$ in the lower level market clearing process, in order to account for price-sensitive market participants. Since consumers and producers have elastic market curves, they will demand and offer different quantities at different prices.

\subsubsection{Upper Level}\label{sec:UpperLevel}

The upper level problem represents a long term investment planning problem, and is formulated as follows: 
\begin{subequations}\label{upperLevel}
\begin{align}
\max_{\substack{\uBinary, \bBinaryFlowMax, \tariff \\ \priceDemand, \priceGen}} &\sum_{t \in \setTime} \discountFactor \big(\sum_{k \in \setDemandAll} \priceDemandTilde\demand - \sum_{p \in \setGenAll} \priceGenTilde\gen \big) \notag\\
&- \sum_{m \in \setLines} (\uBinary\KFix + \KVar\sum_{j \in \setLumpyInvestments} \bBinaryFlowMax\flowMaxDiscrete) \label{upperObjFunction}
\end{align}
\vspace*{-2em}
\begin{align}
&\text{s.t.}\notag\\
&\sum_{j \in \setLumpyInvestments} \bBinaryFlowMax = \uBinary  \hspace{8em}  \forall m \in \setLines \label{upperBFlowSumEqualU} \\
&\priceDemand = \priceDemandTilde - \sum_{m \in \setLines} \uBinary\tariff\PTDF \qquad \forall t \in \setTime, \forall n \in \setNodes, \forall k \in \setDemandNode \label{upperDemandTilde}\\ 
&\priceGen = \priceGenTilde + \sum_{m \in \setLines} \uBinary\tariff\PTDF \qquad  \forall t \in \setTime, , \forall n \in \setNodes, \forall p \in \setGenNode \label{upperGenTilde}\\
&\sum_{t \in \setTime} \sum_{m \in \setLines} \sum_{n \in \setNodes}  - \discountFactor \incidenceMatrixElement \price \flow  \notag\\
&+\sum_{t \in \setTime} \sum_{m \in \setLines} \sum_{n \in \setNodes} \discountFactor \uBinary\tariff\PTDF \big(\sum_{k \in \setDemandNode} \demand + \sum_{p \in \setGenNode} \gen\big) \geq \notag\\
&\sum_{m \in \setLines} (\uBinary\KFix + \KVar\sum_{j \in \setLumpyInvestments}\bBinaryFlowMax\flowMaxDiscrete)\label{upperLevelRevenueAdequacy} \\
&\uBinary \in \{0,1\} , \quad \bBinaryFlowMax \in \{0,1\} , \quad \tariff \geq 0 ,  \quad \priceDemand \in \mathbb{R} , \quad \priceGen \in \mathbb{R}. \label{upperLevelVariables}
\end{align}
\end{subequations}
The objective function \eqref{upperObjFunction} represents the long term social welfare and accounts for both the fixed investment cost $\KFix$ and variable cost $\KVar$. The term $\discountFactor$ is a weighting factor to consider the different time periods involved. The term $\flowMaxDiscrete$ represents the lumpy expansion on line $m$. In the proposed framework, the line capacity can be increased only by a finite set of discretised quantities $\setLumpyCapExpansionsF=\bigcup_{j \in \setLumpyInvestments} \flowMaxDiscrete$. Constraint \eqref{upperBFlowSumEqualU} ensures that only one value of $\flowMaxDiscrete$ is selected if the investment is made. Constraints \eqref{upperDemandTilde}-\eqref{upperGenTilde} introduce the effect of network tariffs into the problem. For the case of demand in \eqref{upperDemandTilde}, 
the actual marginal benefit $\priceDemand$ of the consumer $k$ at time $t$ is reduced by a quantity $\tariff$ (i.e. the imposed tariff) with respect to its initial gross benefit $\priceDemandTilde$. Equivalently, the market demand curve is reshaped to account for the additional costs. A similar formulation is used for generators in \eqref{upperGenTilde}, where $\tau$ appears with a plus sign since costs (rather than benefits) are considered. In this case, the tariffs increase the producer costs, and the market supply curve is reshaped accordingly. The market curves determined by using $\priceDemand$ and $\priceGen$, which embed network costs, will be used by the lower level problem to perform the market clearing. Note that the network tariffs $\tariff$ do not appear in \eqref{upperObjFunction}, because they are paid by consumers and producers and collected by the network operator to recover the investment costs, therefore these monetary flows offset each other in \eqref{upperObjFunction}.
 
 \begin{remark}
 The term $\PTDF$ in \eqref{upperDemandTilde}-\eqref{upperGenTilde} represents an allocation factor to account for possible different network cost distributions between the participants. For example, a postage stamp-like allocation method could be enforced by imposing $\PTDF=1$ $\forall m, n$. Alternatively, network distribution factors could be used to account for the different impact of demand/generation at each node on the individual power flows. Negative values of $\PTDF$ imply that a credit is collected.
 \end{remark}

Constraint \eqref{upperLevelRevenueAdequacy} ensures the revenue adequacy of the network expansion. The terms $- \incidenceMatrixElement \price \flow$ refer to the congestion rent (the minus sign is due to the sign convention of $\incidenceMatrixElement$). The remaining terms in the left-hand-side of \eqref{upperLevelRevenueAdequacy} represent the tariffs collected from consumers and producers, respectively.  
The terms in the right-hand-side are the fixed and variable costs of the network expansion. The decision variables of the upper level problem are $\uBinary$, $\bBinaryFlowMax$, $\tariff$, $\priceDemand$, and $\priceGen$. The terms $\demand$, $\gen$, and $\flow$ are the lower level optimal values, corresponding to the term $x^*$ in \eqref{sketchBilevel}.

\subsubsection{Lower Level}\label{sec:LowerLevel}

The lower level problem is a standard market clearing problem, e.g. day-ahead market, for each time $t \in \setTime$. As long as the clearing problems are independent, a single program spanning over all time periods can be formulated as follows:
\begin{subequations}\label{lowerLevel}
\begin{align}
\max_{\demand, \gen, \flow} & \sum_{t \in \setTime} \sum_{n \in \setNodes} \bigg(\sum_{k \in \setDemandNode} \priceDemand\demand - \sum_{p \in \setGenNode} \priceGen\gen\bigg) \label{lowerLevelObjectiveFunction}
\end{align}
\vspace*{-2em}
\begin{align}
&\text{s.t.}\notag\\
&\demand \leq \demandMax && \forall t \in \setTime, \forall k \in \setDemandAll && [\phiDemMax \geq 0] \label{demandQuantityConstraint}\\
&\gen \leq \genMax && \forall t \in \setTime, \forall p \in \setGenAll && [\phiGenMax \geq 0] \label{generationQuantityConstraint}
\end{align}
\vspace*{-2em}
\begin{align}
&\sum_{k \in \setDemandNode}\demand - \sum_{p \in \setGenNode}\gen + \sum_{m \in \setLines} \incidenceMatrixElement \flow = 0  \notag\\& \hspace{7em} \forall t \in \setTime, \forall n \in \setNodes && [\price \in \mathbb{R}] \label{lowerPowerBalance}
\end{align}
\vspace*{-2em}
\begin{align}
&\sum_{m \in \setLines} \sensitivityMatrixElementKVL \flow = 0 && \forall t \in \setTime, \forall \ell \in \setLoops && [\gammaKVL \in \mathbb{R}]\label{KVLContraint}\\
&\flow \leq \flowZero + \sum_{j \in \setLumpyInvestments} \bBinaryFlowMax\flowMaxDiscrete && \forall t \in \setTime, \forall m \in \setLines && [\muMax \geq 0]\label{flowMaxConstraint}\\
&- \flow \leq \flowZero + \sum_{j \in \setLumpyInvestments} \bBinaryFlowMax\flowMaxDiscrete && \forall t \in \setTime, \forall m \in \setLines && [\muMin \geq 0] \label{flowMinConstraint} \, .
\end{align}
\end{subequations}
Dual variables are reported in square brackets. The objective function \eqref{lowerLevelObjectiveFunction} maximizes the social welfare of the participants, while accounting for network tariffs as determined by \eqref{upperDemandTilde}-\eqref{upperGenTilde}. That is, the clearing process is performed by considering all the costs sustained by participants, in order to properly account for the behaviour of price-elastic consumers and producers \cite{mankiw2012principles, kirschen2004fundamentals}. Constraints \eqref{demandQuantityConstraint}-\eqref{generationQuantityConstraint} impose maximum quantity limits. Condition \eqref{lowerPowerBalance} represents the power balance in node $n$ at time $t$. Note that the dual variable $\price$ associated with  this constraint represents the market clearing price according to the marginal pricing framework \cite{schweppe1988spot}. Constraint \eqref{KVLContraint} enforces the Kirchhoff's voltage law. Finally, constraints \eqref{flowMaxConstraint}-\eqref{flowMinConstraint} determine the flow limits, where $\flowMaxDiscrete$ is the lumpy expansion as determined by the upper level. The decision variables of the lower level problem are $\demand$, $\gen$, and $\flow$. As outlined in Section~\ref{sec:BilevelProgramming}, all the upper level variables enter the lower level as parameters, therefore \eqref{lowerLevel} is a linear program, which is a key property to facilitate resolution of the overall bilevel problem.

\section{Resolution method}\label{sec:resolutionMethod}

Bilevel problems like the one presented in (\ref{upperLevel}) and (\ref{lowerLevel}) are often solved iteratively, considering a sequential separate resolution of their upper and lower level until convergence to a fixed point \cite{dempe2002foundations}. However, this methodology might be time consuming and does not ensure convergence. For this reason, in this section we propose a one-shot resolution method obtained by i) recasting the bilevel problem as a single nonlinear optimization program, ii) introducing a novel and equivalent reformulation of the congestion rent and iii) using standard integer algebra to remove critical nonlinearities. The final result of this analytical process is the MILP presented in \ref{appendixMILP}, which can be solved with off-the-shelf solvers. 

\subsection{Single Level Problem}\label{sec:SingleLevel}

As stated in Section~\ref{sec:LowerLevel} the lower level problem \eqref{lowerLevel} is a linear program. As a consequence, it can be equivalently represented by using its first order necessary and sufficient Karush-Kuhn-Tucker (KKT) conditions \cite{boyd2004convexOptimization}. However, the KKT complementarity slackness conditions are nonlinear relations. In order to avoid these nonlinearities, the strong duality property can be used instead of the KKT complementarity slackness conditions. The analytical details for this procedure are detailed in \ref{appendixDualConstraint}. The key principle is that, for a feasible linear program, the strong duality property holds if and only if all the complementarity slackness conditions hold \cite{bradley1977applied}. The strong duality requires the equivalence between the primal and dual objective function values at the optimum, and is reported in \eqref{strongDuality}. As a consequence, the proposed bilevel model can be recast as a single level problem, where the lower level is represented into the upper level by using its primal constraints \eqref{demandQuantityConstraint}-\eqref{flowMinConstraint}, its dual constraints \eqref{dualDemandQuantity}-\eqref{dualFlowConstraint}, and the strong duality property \eqref{strongDuality}. To summarize, the bilevel model described in Section \ref{sec:UpperLevel} and Section \ref{sec:LowerLevel} can be equivalently recast as a single mathematical optimization program, defined as follows:
\begin{subequations}\label{singleLevel}
\begin{align}
\max \quad &\sum_{t \in \setTime} \discountFactor \big(\sum_{k \in \setDemandAll} \priceDemandTilde\demand - \sum_{p \in \setGenAll} \priceGenTilde\gen \big) \notag\\
& - \sum_{m \in \setLines} (\uBinary\KFix + \KVar\sum_{j \in \setLumpyInvestments} \bBinaryFlowMax\flowMaxDiscrete) \label{singleObj}\\
&\eqref{upperBFlowSumEqualU}-\eqref{upperLevelVariables} \label{singleUpperLevel}\\
&\eqref{demandQuantityConstraint}-\eqref{flowMinConstraint}\label{singleLowerLevelPrimal}\\
&\eqref{dualDemandQuantity}-\eqref{dualFlowConstraint} \label{singleLowerLevelDual}\\
&\eqref{strongDuality} \label{singleStrongDuality}
\end{align}
\end{subequations}
where \eqref{singleUpperLevel} represents the upper level constraints, \eqref{singleLowerLevelPrimal} and \eqref{singleLowerLevelDual} refer to the lower level primal and dual constraints, and \eqref{singleStrongDuality} is the strong duality property. The decision variables of the single level problem are the upper level variables, and both the primal and dual lower level variables, i.e., $\uBinary$, $\bBinaryFlowMax$, $\tariff$, $\priceDemand$, $\priceGen$, $\demand$, $\gen$, $\flow$, $\phiDemMax$, $\phiGenMax$, $\price$, $\gammaKVL$, $\muMax$, $\muMin$. Note that, by recasting the bilevel problem as a single optimization problem, both lower level primal and dual variables are accessible within the same optimization problem. This key design feature is exploited to compute the transmission system operator revenues in \eqref{upperLevelRevenueAdequacy}. The single level problem \eqref{singleLevel} is a nonlinear integer program, Section~\ref{sec:CongestionRentRecast} and Section~\ref{sec:MILP} show how all the nonlinearities can be removed.

\subsection{Congestion rent recast}\label{sec:CongestionRentRecast}

This section presents an equivalent characterization of the congestion rent, derived through optimality conditions. This novel result allows us to recast the congestion rent as a linear expression in the presence of lumpy investments. The key feature of the proposed approach is the use of complementarity conditions to derive an expression of the congestion rent that does not depend on the flow terms $\flow$, which is a key point. This property will be exploited to recast the whole bilevel problem as a single MILP program. Indeed, the congestion rent collected by the transmission system operator is defined as follows:
\begin{equation}\label{congestionNonLinear}
\sum_{m \in \setLines} \sum_{n \in \setNodes} - \incidenceMatrixElement \price \flow\, ,
\end{equation}
which is a nonlinear relation, as it involves the products of market prices $\price$ and flows $\flow$, i.e. two continuous variables. However, by exploiting complementarity conditions, it is possible to recast \eqref{congestionNonLinear} as follows (see \ref{appendixCongestionRecast}):
\begin{equation}
\sum_{m \in \setLines} (\flowZero + \sum_{j \in \setLumpyInvestments} \bBinaryFlowMax\flowMaxDiscrete)(\muMax + \muMin) \label{congestionRentRecast}
\end{equation}
that can be further rewritten as a linear expression when lumpy expansions are considered, as described in Section \ref{sec:MILP}. By using \eqref{congestionRentRecast}, the revenue adequacy constraint \eqref{upperLevelRevenueAdequacy} can be recast as follows:
\begin{align}\label{congestionRentRecastFinal}
&\sum_{t \in \setTime}  \sum_{m \in \setLines} \discountFactor(\flowZero + \sum_{j \in \setLumpyInvestments} \bBinaryFlowMax\flowMaxDiscrete)(\muMax + \muMin)  \notag\\
&+\sum_{t \in \setTime} \sum_{m \in \setLines} \sum_{n \in \setNodes} \discountFactor \uBinary\tariff\PTDF \big(\sum_{k \in \setDemandNode} \demand + \sum_{p \in \setGenNode} \gen\big) \geq \notag\\
&\sum_{m \in \setLines} (\uBinary\KFix + \KVar\sum_{j \in \setLumpyInvestments} \bBinaryFlowMax\flowMaxDiscrete)
\end{align}

\subsection{MILP model}\label{sec:MILP}

By using \eqref{congestionRentRecastFinal} instead of \eqref{upperLevelRevenueAdequacy}, only two types of nonlinearities remain in the single level problem \eqref{singleLevel}:
\begin{enumerate}
	\item the products $\bBinaryFlowMax\muMax$ and $\bBinaryFlowMax\muMin$ involving the binary variable $\bBinaryFlowMax$ and the continuous variables $\muMax$ and $\muMin$, in \eqref{congestionRentRecastFinal} and \eqref{strongDuality}.
	\item the products $\uBinary\tariff\demand$ and $\uBinary\tariff\gen$ involving the binary variable $\uBinary$ and the continuous variables $\tariff$, $\demand$, and $\gen$, in \eqref{congestionRentRecastFinal} and \eqref{strongDuality}. These nonlinearities appear in \eqref{strongDuality} when the conditions $\eqref{upperDemandTilde}$ and $\eqref{upperGenTilde}$ are used to replace the terms $\priceDemand$ and $\priceGen$. 
\end{enumerate}

The nonlinearities of the first type involve only only the product between one binary variable and one continuous variable, and can be linearized exactly by using standard integer algebra. For example, the product $bx$ of the binary variable $b$, and the continuous variable $x$ with bounds $\pm M$, can be equivalently replaced by introducing an auxiliary continuous bounded variable $y$ defined as follows:
\begin{align}
&-Mb \leq y \leq +Mb\label{aux1}\\
&-M(1-b) \leq x - y \leq +M(1-b)\label{aux2} \, .
\end{align}
Therefore, by exploiting \eqref{aux1}-\eqref{aux2}, the nonlinear terms\\ $\bBinaryFlowMax \muMax$ and $\bBinaryFlowMax\muMin$ are removed by introducing the auxiliary variables $\auxBFMuMaxU$ and $\auxBFMuMinU$, defined in \eqref{auxConstraintBF1}-\eqref{auxConstraintBF2}.\\

To linearize the second type of nonlinearities, discrete tariff levels are assumed. As a consequence, the term $\uBinary\tariff$ is replaced by the term $\sum_{i \in \setTariff} \bBinaryTau \tariffDiscrete$, with $\tariffDiscrete \in \mathbb{R}$, where $i \in \setTariff$ represents the number of possible different tariffs for each line $m$, with:
\begin{align}
&\sum_{i \in \setTariff} \bBinaryTau = \uBinary  && \forall m \in \setLines\label{conditionsBTau1}\\
&\bBinaryTau \in \{0, 1\} && \forall m \in \setLines, \forall i \in \setTariff \label{conditionsBTau2} \, .
\end{align}
Therefore, the term $\uBinary\tariff\demand$ and $\uBinary\tariff\gen$ are replaced by $\sum_{i \in \setTariff}\bBinaryTau \tariffDiscrete\demand$, and $\sum_{i \in \setTariff}\bBinaryTau \tariffDiscrete\gen$, respectively. Then, the product $\bBinaryTau\demand$ and $\bBinaryTau\gen$, involving one binary and one continuous variable, is removed as in \eqref{aux1}-\eqref{aux2} by  introducing the auxiliary variables $\auxBTauDemand$ and $\auxBTauGen$  defined in \eqref{auxConstraintBTauDemand1}-\eqref{auxConstraintBTauGen2}. \\

Assuming discrete tariff levels, by using \eqref{congestionRentRecastFinal} instead of \eqref{upperLevelRevenueAdequacy}, and introducing the constraints \eqref{auxConstraintBF1}-\eqref{auxConstraintBTauGen2}, all the nonlinearities of the single level program \eqref{singleLevel} can be removed, and the final optimization problem results in a MILP model, fully reported in \ref{appendixMILP}, which can be solved with off-the-shelf solvers.

\section{Numerical results}\label{sec:numericalResults}
The new network planning formulation with ex-ante tariffs is now applied to two different case studies, in order to assess and discuss its key properties and advantages. The proposed approach is also compared with other models referenced in the paper. In particular, the following planning approaches are considered:
\begin{itemize}
	\item \textbf{Centralized Solution} (CS): the standard centralized planning approach, presented in \eqref{CS}.
	\item \textbf{Centralized Solution with revenue adequacy} (CSR): extension of CS where transmission costs must be fully recovered from congestion rent. This is described by equations \eqref{CS} with the additional constraint \eqref{revenueAdequacyIntro}.
	\item{\textbf{Centralized Solution with revenue adequacy and lumpy expansion} (CSR-L)}: modification of CSR that considers lumpy investment decisions. In this case, the capacity expansion $\flowMax$ on each line $m$ can only take a finite set of values in $\setLumpyCapExpansionsF$.  Therefore, it is described by \eqref{CS}, \eqref{revenueAdequacyIntro} and the following additional constraint:
	\begin{equation}
	\flowMax \in \setLumpyCapExpansionsF, \qquad \forall \, m \in \mathcal{M}.
	\end{equation}
	\item \textbf{Tariff Solution} (TS): the new planning paradigm with ex-ante dynamic tariffs presented in this paper, which is fully reported in \ref{appendixMILP}.
\end{itemize}

In the reported studies, the MILP model of the TS has been implemented in Python 3.6 with Pyomo 5.6 \cite{pyomoBook}, and solved with CPLEX 12.8 \cite{cplex2009v12} on a 16-core Intel Xeon CPU, with 32GB of RAM. To guarantee the optimality of the solution, the absolute and relative Cplex gaps are set to zero.

\subsection{2-node system}\label{numeric:2node}
\begin{table*}[ht]
	\boldmath
	\centering
	\setlength{\extrarowheight}{1pt}
	\caption{Results of different network expansion schemes applied to the 2-node system.}
	\label{tab:increaseW}
	\begin{tabular}{|c|r|r|r|r|r|r|r|}
		\hline
		\multicolumn{1}{|c|}{\textbf{\begin{tabular}[c]{@{}c@{}}Reference\\model\end{tabular}}} & \multicolumn{1}{c|}{\textbf{\begin{tabular}[c]{@{}c@{}}Expansion \\ $\flowMax$ (or $\flowMaxDiscrete$) \end{tabular}}} & \multicolumn{1}{c|}{\textbf{\begin{tabular}[c]{@{}c@{}}Social Welfare\\ increase\end{tabular}}} & \multicolumn{1}{c|}{\textbf{\begin{tabular}[c]{@{}c@{}}Investment\\Cost\end{tabular}}} & \multicolumn{1}{c|}{\textbf{\begin{tabular}[c]{@{}c@{}}Congestion\\ Rent\end{tabular}}} & \multicolumn{1}{c|}{\textbf{\begin{tabular}[c]{@{}c@{}}Tariff \\ $\tariff$ \end{tabular}}} & \multicolumn{1}{c|}{\textbf{\begin{tabular}[c]{@{}c@{}}Tariff \\ Payments\end{tabular}}} & \multicolumn{1}{c|}{\textbf{\begin{tabular}[c]{@{}c@{}}Revenue\\imbalance\end{tabular}}} \\ \hline
		\textbf{CS}&		25                      & 425                                                                                & 450                                                                                      & 250                                                                                          & 0                                       & 0                                                                                              & -200                                                                                                    \\ \hline
		\textbf{CSR} &		20                      & 400                                                                                & 400                                                                                      & 400                                                                                          & 0                                       & 0                                                                                              & 0                                                                                                       \\ \hline
		\textbf{CSR-L}&		18                      & 376                                                                                & 380                                                                                      & 432                                                                                          & 0                                       & 0                                                                                              & 52                                                                                                      \\ \hline
		\textbf{TS}&		21                      & 409                                                                                & 410                                                                                      & 371                                                                                          & 0.357                                    & 39                                                                                             & 0                                                                                                       \\ \hline
	\end{tabular}
\end{table*}
The 2-node network represented in Fig.~\ref{fig:2node} and analysed in Section \ref{sec:fixedcosts} is now further discussed. Recall that the fixed cost is assumed equal to $\KFix=200$~\pounds/h and the variable cost is set to $\KVar=10$~\pounds/MWh. The set of lumpy capacity expansions used for both the CSR-L and the TS model is $\setLumpyCapExpansionsF =\{3, 6, 9, \ldots\, 30\}$. The demand and supply profiles are fully reported in Fig.~\ref{fig:zone12}. Table \ref{tab:increaseW} reports the solutions associated to each of the four considered investment schemes, i.e. CS, CSR, CSR-L, and TS.
The CS proposes investment on a line of capacity $\flowMax=25$ MW, as graphically represented in Fig. \ref{fig:IDES}. This centralized approach leads to a net increase of $425$ \pounds/h in the social welfare of the system with respect to the initial no-line scenario. The total investment cost required to achieve this result is equal to $450$ \pounds/h, which can be disaggregated into the variable cost $\KVar \flowMax= 250$ \pounds/h and fixed cost $\KFix=200$ \pounds/h. 
The proper recovery and allocation of this cost remains an open problem. As discussed in Section \ref{sec:fixedcosts}, the congestion rent CR in the expanded network is equal to $250$ \pounds/h and only covers variable costs, causing a revenue imbalance of $-200$ \pounds/h. One possibility to recover this remaining cost term is the introduction of ex-post tariffs, i.e. additional payments to be made by demand and generators in exchange for benefiting of the new line. However, as discussed in Section \ref{sec:expost_tariffs}, this can lead to a negative surplus for some participants. In addition, it disincentives truthful biddings, as shown in \cite{bushnell2014EfficiencyTrasmissionCost}. In other words, the CS approach, with social welfare increase of $425$~\pounds/h, does not properly address the key issue of full cost recovery for the network expansion.

The alternative approach considered in CSR is to explicitly impose full cost recovery by using the collected congestion rent, through the additional constraint \eqref{revenueAdequacyIntro}. Comparison between CS and CSR shows the following:
\begin{itemize}
	\item A smaller expansion of $\flowMax=20$ MW is performed in the CSR case, as represented in Fig. \ref{fig:netIDES2}.
	\item The underinvestment leads to a higher price differential between the two areas.
	\item This increases the congestion rent to $400$ \pounds/h, which now fully covers the investment costs, i.e. the variable cost $\KVar \flowMax= 200$ \pounds/h plus the fixed cost $\KFix=200$ \pounds/h.
	\item The underinvestment reduces the net social welfare increase, which in the CSR is equal to $400$ \pounds/h (compared to $425$ \pounds/h in the CS case).
\end{itemize} 
To summarize, the CRS is able to ensure full cost recovery by using exclusively the collected congestion rent. However, this is achieved at the cost of a reduced social welfare.

The drawbacks of the CRS are exacerbated in the CRS-L scenario, when lumpy investments are considered and the capacity expansion of the line can only take a finite number of values. In this case, the line expansion is smaller and equal to $\flowMax=18$~MW (see Fig. \ref{fig:Lumpy}), since the value of $20$~MW of the CSR case cannot be selected. This implies that the collected congestion rent increase even further to $432$ \pounds/h, leading to a positive revenue imbalance of $+52$ \pounds/h (red area in Fig.~\ref{fig:Lumpy}). As expected, the CRS-L has a smaller net benefit increase of the overall social welfare, which amounts to only $376$ \pounds/h.

Finally, the last row in Table~\ref{tab:increaseW} reports the solution obtained by using the proposed TS approach. It involves 203 binary variables and is solved in 1107.40 seconds. In this case, the selected network expansion $\flowMax=21$ MW is greater than the one in both the CSR and CSR-L case. As a result, the smaller congestion rent of $371$ \pounds/h does not fully cover the investment costs of $410$ \pounds/h. The remaining difference of $39$ \pounds/h is obtained by applying a tariff $\tau=0.357$ £/MWh. This corresponds to a price increase in demand payments (and a price decrease for generation revenues). In other words, demand and generation sustain an additional cost in exchange for the (greater) benefits of the expanded line. A key advantage of this approach is the greater net increase in social welfare (equal to $409$\pounds/h) with respect to CSR and CSR-L. This is obtained by introducing an additional degree of freedom in the problem of recovering investment costs: through the introduction of the tariff $\tau$, these costs are not entirely covered by the congestion rent but can in part be sustained by the agents (demand and generation) which benefit from the investment. Notice that the distribution of the cost recovery (between congestion rent and demand/generation payments) is determined by $\tau$, which is chosen with the purpose of maximizing the overall social welfare in (\ref{upperObjFunction}).

For comparison, the results with \textit{ex-post} tariff can be obtained by using the CS model, and imposing the expansion of $21$ MW. The key difference between the two outcomes is that in the ex-ante solution the quantity cleared is smaller by $0.53$~MWh. This quantity belongs to participants that suffer a loss when the tariff is applied ex-post, and quit the market when the tariff is applied ex-ante.

\subsection{Garver's 6-node system}\label{numeric:6-node}
\begin{figure}[h!]
	\centering
	\includegraphics[width=0.45\textwidth]{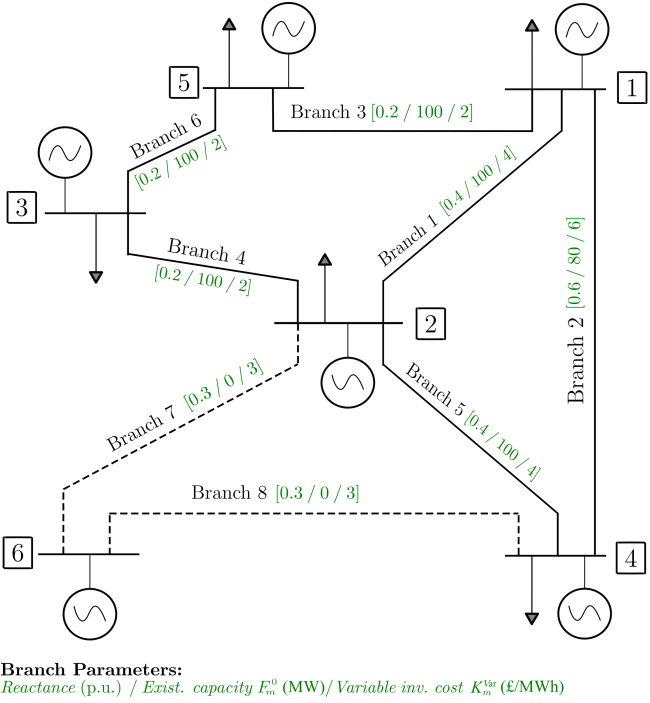}
	\caption{Garver's 6-node network topology.}
	\label{fig:garver}
\end{figure}

\begin{table*}[ht!]
	\boldmath
	\centering
	\setlength{\extrarowheight}{1pt}
	\caption{Results of different network expansion schemes applied to the Garver's 6-node system.}
	\label{tab:6bus}
	\begin{tabular}{|c|r|r|r|r|r|r|r|r|r|}
		\hline
		\multirow{2}{*}{\textbf{\begin{tabular}[c]{@{}c@{}}Reference\\ Model\end{tabular}}} & \multicolumn{2}{c|}{\textbf{Expansion}}           & \multicolumn{1}{c|}{\multirow{2}{*}{\textbf{\begin{tabular}[c]{@{}c@{}}Social \\ Welfare\end{tabular}}}} & \multicolumn{1}{c|}{\multirow{2}{*}{\textbf{\begin{tabular}[c]{@{}c@{}}Investment\\ Cost\end{tabular}}}} & \multicolumn{1}{c|}{\multirow{2}{*}{\textbf{\begin{tabular}[c]{@{}c@{}}Congestion\\ Rent\end{tabular}}}} & \multicolumn{2}{c|}{\textbf{Tariff}}              & \multicolumn{1}{c|}{\multirow{2}{*}{\textbf{\begin{tabular}[c]{@{}c@{}}Tot. Tariff\\ Payments\end{tabular}}}} & \multicolumn{1}{c|}{\multirow{2}{*}{\textbf{\begin{tabular}[c]{@{}c@{}}Revenue\\ Imbalance\end{tabular}}}} \\ \cline{2-3} \cline{7-8}
		& \multicolumn{1}{c|}{$F_7$} & \multicolumn{1}{c|}{$F_8$} & \multicolumn{1}{c|}{}                                                                                    & \multicolumn{1}{c|}{}                                                                                    & \multicolumn{1}{c|}{}                                                                                    & \multicolumn{1}{c|}{$\tau_7$} & \multicolumn{1}{c|}{$\tau_8$} & \multicolumn{1}{c|}{}                                                                                         & \multicolumn{1}{c|}{}                                                                                      \\ \hline
		\textbf{CS}                                                                         & 55.51                   & 44.27                   & 21774.38                                                                                                 & 499.3                                                                                                    & 299.3                                                                                                    & 0                       & 0                       & 0                                                                                                             & -200                                                                                                       \\ \hline
		\textbf{CSR}                                                                        & 50.15                   & 39.91                   & 21762.54                                                                                                 & 470.2                                                                                                    & 470.2                                                                                                    & 0                       & 0                       & 0                                                                                                             & 0                                                                                                          \\ \hline
		\textbf{CSR-L}                                                                      & 50.00                   & 40.00                   & 21762.39                                                                                                 & 470.0                                                                                                    & 471.6                                                                                                    & 0                       & 0                       & 0                                                                                                             & 1.6                                                                                                        \\ \hline
		\textbf{TS}                                                                         & 53.00                   & 48.00                   & 21765.11                                                                                                 & 503.0                                                                                                    & 275.2                                                                                                    & 0.0419                  & 0.0419                  & 227.8                                                                                                         & 0                                                                                                          \\ \hline
	\end{tabular}
\end{table*}
The comparison of the different investment strategies is now extended to a more complex network with a significantly larger number of generators and consumers. The case study is based on the Garver's $6$-node system  depicted in Fig.~\ref{fig:garver} \cite{garver1970}. This network has $6$ nodes and $8$ lines, i.e. $\setNodes=\{1, \ldots, 6\}$ and $\setLines = \{1, \ldots, 8\}$. The solid lines from $1$ to $6$ in Fig.~\ref{fig:garver} represent existing branches that can be expanded, i.e. $\flowZero > 0$ for all $m \in \{1, \ldots, 6\}$. The dashed lines $7$ and $8$ are new branches that can be built, i.e. $\flowZero = 0$ for $m \in \{7, 8\}$. 
In this example, two time periods $\setTime=\{1,2\}$ are considered. To properly depict elastic market curves, $1000$ consumers and $1000$ generators are considered at each node, with the exception of node $6$, where only generators are present. The values of $\priceDemandTilde$ and $\priceGenTilde$ are sampled from a normal distribution with mean $50$~\pounds/MWh and standard deviation $10$~\pounds/MWh. The limits $\demandMax$ and $\genMax$ on demand and supply bids are obtained from an uniform distribution ranging from zero to $500$ kW. The set of lumpy capacity expansions is defined as $\bar{F}_m \in \{1, 2, 3,  \ldots, 200\}$. The line reactance, existing capacity $\flowZero$, and variable cost $\KVar$ are reported for each branch in Fig.~\ref{fig:garver}. The fixed investment cost is $\KFix=100$~\pounds/h for all $m \in \setLines$.

Table~\ref{tab:6bus} reports the solutions associated to each of the four considered schemes (i.e., CS, CSR, CSR-L and TS).  The optimal solution obtained by using the CS scheme shows that only the lines $7$ and $8$ are built, with $\bar{F}_7=55.51$ MW and $\bar{F}_8=44.27$ MW, leading to a welfare of $21774.38$~\pounds/h. The total investment cost amounts to $499.3$~\pounds/h, given by the fixed investment cost of $\KFix=100$~\pounds/h and $\KVar=3$~\pounds/MWh for both the lines. In this case, the congestion rent is $299.3$~\pounds/h and, as expected, it exactly covers the variable investment costs, but not the fixed costs. As a result, the revenue imbalance is negative, and equal to $-200$~\pounds/h. In order to ensure the recovery of all the investment costs, the CRS scheme with the additional revenue adequacy constraint \eqref{revenueAdequacyIntro} is considered, with the associated solution reported in the second row of Table~\ref{tab:6bus}. In this case, as one would expect, the line expansions are smaller: line $7$ is built with $\bar{F}_7=50.15$ MW, whereas line $8$ is built with $\bar{F}_8=39.91$ MW. As a result, the welfare decreases to 21762.54~\pounds. However, due to the underinvestment, the congestion rent rises to $470.2$~\pounds/h, and exactly covers all the investment costs. When the set of lumpy capacity expansions $\bar{F}_m \in \{1, 2, 3,  \ldots, 200\}$ is considered in the CSR-L scheme, the social welfare decreases even further. In this case, line $7$ is built with $\bar{F}_7=50$ MW, whereas line $8$ is built with $\bar{F}_8=40$ MW. As a result of lumpiness, a lower value of social welfare is obtained (21762.39~\pounds) and extra-profit is collected by the transmission operator, leading to a strictly positive revenue imbalance of $1.6$~\pounds/h. 

The last row in Table~\ref{tab:6bus} reports the solution associated to the proposed scheme TS with ex-ante network tariffs. It involves 1696 binary variables and is solved in 4458.26 seconds. In this case, line $7$ is built with $\bar{F}_7=53$ MW, whereas line $8$ is built with $\bar{F}_8=48$ MW. Three main elements should be emphasized in this case: 
\begin{itemize}
	\item The social welfare is equal to $21765.11$~\pounds/h and it is greater than both the CSR-L and CSR cases.
	\item By imposing an ex-ante tariff equal to $0.0419$~\pounds/MWh and directly recovering part of the investment costs from generators and customers, the TS avoids the negative revenue imbalance arising in the CS scenario. 
	\item The CSR and the TS have both a revenue imbalance (i.e. the surplus of the transmission operator) equal to zero. This means that, in the TS, the increase in social welfare with respect to the CRS case is entirely collected by consumers and producers. In other words, despite the additional tariff payment, the condition of the market participants improves thanks to the optimal distribution of the investment cost recovery between the tariff and congestion rent component.
\end{itemize}
	
\subsection{Investment and economic trends}
From the numerical case studies presented above, it is possible to draw some general comparisons between the planning solutions and emphasize the advantages of the proposed ex-ante dynamic tariffs. Regarding the Centralized Solution (CS), it can be seen that it achieves the highest improvement in overall social welfare, both in the 2-bus and 6-bus networks. However, such improvement is obtained at the expense of a negative revenue imbalance. The overall social welfare is reduced when the full recovery of costs is included as a constraint in the CSR, as underinvestment is performed to lower costs and increase congestion revenues. For example, the social welfare of the CSR in the 2-bus network is 5.6\% lower than in the CS case. Similarly, in the Garver system, the additional social welfare of the network expansion with the CSR is 0.05\% lower. This trend is even more evident when lumpy expansions are considered in the CSR-L solution.

Within this framework, the new proposed solution TS achieves significant improvements. Thanks to the introduction of dynamic network tariffs, only a fraction of the investment cost needs to be recovered through congestion revenues (the 90\% and the 54.7\% in the 2-node and 6-node network, respectively). This means that network costs can be fully recovered without necessarily resorting to underinvestment, which leads to higher social welfare. With respect to the CSR-L case, the network investments in the TS are greater by 7.9\% in the 2-node network and by 7\% in the Garver system. As a result, the social welfare metric is improved by 8.7\% and 0.01 \% in the 2-node and 6-node network, respectively.

\section{Conclusion}\label{sec:conclusion}
Current paradigms for the recovery of investment costs in transmission network planning exhibit substantial issues. In particular, the use of ex-post network tariffs can lead to several inefficiencies in the form of cleared orders with a negative surplus and mistrustful biddings. These problems are also exacerbated by the current trends of increasing price-elasticity of demand. To overcome these issues, the proposed approach introduces ex-ante dynamic network tariffs, which can be exactly embodied in the market participants' submitted bids before the market clearing actually takes place, and are directly related to the cleared quantity. These ex-ante tariffs are incorporated in a novel formulation of the long-term network planning problem which returns the optimized network investments and, at the same time, determines the optimal repartition of the associated costs between congestion revenues and tariff payments. 

The paper demonstrates that ex-ante dynamic network tariffs can be used effectively to account for the increasing price-elasticity of demand in the network planning problem while avoiding undesired phenomena such as mistrustful bidding and negative surplus for generators and consumers.  
Through the application of standard linear algebra, complementarity properties and a reformulation of the congestion rent, it has been 	
shown that the network planning problem with ex-ante tariffs can be formulated as a standard mixed-integer linear problem. This ensures optimality while preserving key features of the transmission expansion decisions, such as lumpiness of the investments. Finally and most importantly, it has been demonstrated with relevant case studies that the use of ex-ante network tariffs i) ensures full recovery of the investment costs and ii) guarantees an increased efficiency of the network investments and an improvement of the social welfare when price-elastic demand is considered. 

Further work will investigate more complex formulations of the ex-ante tariffs, evaluating how different allocation factors and tariff specifications affect the optimal network expansion and the fairness of cost repartition. In particular, a comparison with existing tariff allocation approaches, such as the flow-based MW-mile method, is the subject of an ongoing companion paper. More realistic and comprehensive models will also be considered. Future research will move from the current deterministic approach to a stochastic model which takes into account key factors such as wind availability, system load and demand price-sensitivity, whose evolutions on long time scales are not perfectly known a priori. The analysis will also incorporate relevant operational constraints such as ramp rates and minimum up/down time. Finally, the modelling framework will be expanded to also consider generation expansion decisions and capture the physical and economical interactions between generation and transmission planning.

\appendix

\section{Dual constraints and strong duality}\label{appendixDualConstraint}

The dual constraints of the lower level problem \eqref{lowerLevel} are defined as follows. In particular, the dual constraints of the primal variables $\demand \geq 0$ and $\gen \geq 0$ are, respectively:
\begin{align}
&\phiDemMax + \price \geq \priceDemand && \forall t \in \setTime, \forall n \in \setNodes, \forall k \in \setDemandNode \label{dualDemandQuantity} \\
&\phiGenMax - \price \geq - \priceGen && \forall t \in \setTime, \forall n \in \setNodes, \forall p \in \setGenNode \, , \label{dualGenQuantity}
\end{align}
The dual constraint associated with the flow variable $\flow \in \mathbb{R}$ is defined as:
\begin{align}
&\sum_{n \in \setNodes} \incidenceMatrixElement \price + \sum_{\ell \in \setLoops} \sensitivityMatrixElementKVL \gammaKVL + \muMax - \muMin = 0 \notag\\
&\hspace{14em} \forall t \in \setTime, \forall m \in \setLines \, . \label{dualFlowConstraint}
\end{align}

The strong duality property of the lower level problem \eqref{lowerLevel} requires the equivalence between the primal and dual objective function values, and is defined as follows:
\begin{align}
&\sum_{t \in \setTime} \sum_{n \in \setNodes} \bigg(\sum_{k \in \setDemandNode} \priceDemand\demand - \sum_{p \in \setGenNode} \priceGen\gen\bigg) = \notag \\
&\sum_{t \in \setTime} \sum_{k \in \setDemandAll} \phiDemMax \demandMax 
+ \sum_{t \in \setTime} \sum_{p \in \setGenAll} \phiGenMax \genMax
 \notag \\
&+\sum_{t \in \setTime} \sum_{m \in \setLines} \bigg( \flowZero + \sum_{j \in \setLumpyInvestments} \bBinaryFlowMax\flowMaxDiscrete \bigg)\bigg(\muMax + \muMin\bigg)\label{strongDuality}
\end{align}

\section{Auxiliary constraints}\label{appendixAuxiliaryConstraints}

The following constraints are used to define to the auxiliary variables $\auxBFMuMaxU$ and $\auxBFMuMinU$
\begin{align}
&0 \leq \auxBFMuMaxU \leq M\bBinaryFlowMax  & \forall t \in \setTime, \forall m \in \setLines, \forall j \in \setLumpyInvestments \label{auxConstraintBF1}
\end{align}
\vspace*{-1.5em}
\begin{align}
&0 \leq \muMax - \auxBFMuMaxU \leq M(1-\bBinaryFlowMax) \notag \\
&\hspace{11em} \forall t \in \setTime, \forall m \in \setLines, \forall j \in \setLumpyInvestments
\end{align}
\vspace*{-1.5em}
\begin{align}
&0 \leq \auxBFMuMinU \leq M\bBinaryFlowMax & \forall t \in \setTime, \forall m \in \setLines, \forall j \in \setLumpyInvestments
\end{align}
\vspace*{-1.5em}
\begin{align}
&0 \leq \muMin - \auxBFMuMinU \leq M(1-\bBinaryFlowMax)  \notag \\
&\hspace{11em} \forall t \in \setTime, \forall m \in \setLines, \forall j \in \setLumpyInvestments \, , \label{auxConstraintBF2}
\end{align}
where $M$ is set equal to the maximum allowed bid price.\\

The following constraints are used to define to the auxiliary variables $\auxBTauDemand$ and $\auxBTauGen$
\begin{align}
&0 \leq \auxBTauDemand \leq \demandMax\bBinaryTau  && \forall t \in \setTime, \forall m \in \setLines, \forall i \in \setTariff, \forall k \in \setDemandAll \label{auxConstraintBTauDemand1} 
\end{align}
\vspace*{-1.5em}
\begin{align}
&0 \leq \demand - \auxBTauDemand \leq \demandMax(1-\bBinaryTau) \notag \\
&\hspace{8em} \forall t \in \setTime, \forall m \in \setLines, \forall i \in \setTariff, \forall k \in \setDemandAll
\end{align}
\vspace*{-1.5em}
\begin{align}
&0 \leq \auxBTauGen\leq \genMax\bBinaryTau  && \forall t \in \setTime, \forall m \in \setLines, \forall i \in \setTariff, \forall k \in \setGenAll
\end{align}
\vspace*{-1.5em}
\begin{align}
&0 \leq \gen - \auxBTauGen \leq \genMax(1-\bBinaryTau) \notag \\
&\hspace{11em} \forall t \in \setTime, \forall m \in \setLines, \forall i \in \setTariff, \forall k \in \setGenAll \label{auxConstraintBTauGen2}
\end{align}

\section{Congestion Recast}\label{appendixCongestionRecast}

This section shows how the congestion rent \eqref{congestionNonLinear} can be recast as \eqref{congestionRentRecast}. In detail, from the dual constraint \eqref{dualFlowConstraint}, the following relation can be obtained:
\begin{align}
- \sum_{n \in \setNodes} \incidenceMatrixElement \price = \sum_{\ell \in \setLoops} \sensitivityMatrixElementKVL \gammaKVL + \muMax - \muMin  \, .\label{priceDifference}
\end{align}
Furthermore, the strong duality property, used to obtain the single level problem in Section~\ref{sec:SingleLevel}, guarantees that all of the complementarity slackness conditions hold at the optimum. Therefore, any of these conditions can be used. In particular, the complementarity slackness conditions related to constraints \eqref{flowMaxConstraint} and  \eqref{flowMinConstraint} are defined as follows:
\begin{align}
\flow\muMax = (\flowZero + \sum_{j \in \setLumpyInvestments} \bBinaryFlowMax\flowMaxDiscrete)\muMax \label{complementarity1}\\
-\flow\muMin = (\flowZero + \sum_{j \in \setLumpyInvestments} \bBinaryFlowMax\flowMaxDiscrete)\muMin \label{complementarity2}
\end{align}
Therefore, the congestion rent \eqref{congestionNonLinear} can be recast as follows:
\begin{align}
&\sum_{m \in \setLines} \sum_{n \in \setNodes} - \incidenceMatrixElement \price \flow \stackrel{\eqref{priceDifference}}{=}
\notag\\
&\sum_{m \in \setLines} \sum_{\ell \in \setLoops} \sensitivityMatrixElementKVL \gammaKVL \flow +  \sum_{m \in \setLines} (\muMax - \muMin) \flow
\stackrel{\eqref{complementarity1}-\eqref{complementarity2}}{=}\notag\\
&\sum_{m \in \setLines} \sum_{\ell \in \setLoops} \sensitivityMatrixElementKVL \gammaKVL \flow  \notag\\
&+ \sum_{m \in \setLines} (\flowZero + \sum_{j \in \setLumpyInvestments} \bBinaryFlowMax\flowMaxDiscrete)(\muMax + \muMin) = \notag\\
&\sum_{\ell \in \setLoops} \gammaKVL (\sum_{m \in \setLines} \sensitivityMatrixElementKVL \flow)  \notag\\ 
&+ \sum_{m \in \setLines} (\flowZero + \sum_{j \in \setLumpyInvestments} \bBinaryFlowMax\flowMaxDiscrete)(\muMax + \muMin) 
\stackrel{\eqref{KVLContraint}}{=}\notag\\
&\sum_{m \in \setLines} (\flowZero + \sum_{j \in \setLumpyInvestments} \bBinaryFlowMax\flowMaxDiscrete)(\muMax + \muMin) \label{congestionRentRecast2} \, ,
\end{align}
where the first equality is due to \eqref{priceDifference}, the second equality is due to the complementarity conditions \eqref{complementarity1}-\eqref{complementarity2}, whereas the last equality is due to the Kirchhoff's voltage law constraint \eqref{KVLContraint}. 
Note that the condition \eqref{KVLContraint} holds for any loop $\ell$. This means that the equivalence obtained in \eqref{congestionRentRecast2} holds not only at system level, but also for any loop $\ell$ at time $t$. Furthermore, it involves only the products of the binary variable $\bBinaryFlowMax$ and the continuous variables $\muMax$ and $\muMin$, which can be linearized using standard integer algebra, as described in the Section~\ref{sec:MILP}. \\

\section{Final MILP model}\label{appendixMILP}

The following appendix reports the complete MILP model after that all the nonlinearities have been removed.
\begin{align}
\max \quad &\sum_{t \in \setTime} \discountFactor\Big(\sum_{k \in \setDemandAll} \priceDemandTilde\demand - \sum_{p \in \setGenAll} \priceGenTilde\gen\Big) \notag\\
&- \sum_{m \in \setLines} \Big(\uBinary\KFix + \KVar\sum_{j \in \setLumpyInvestments}\bBinaryFlowMax\flowMaxDiscrete\Big) \label{AppendixMILPObjFunc}\\
\text{s.t.}&\notag\\
&\eqref{upperBFlowSumEqualU}\\
&\sum_{t \in \setTime} \sum_{m \in \setLines} \discountFactor\flowZero(\muMax + \muMin) \notag\\
& + \sum_{t \in \setTime} \sum_{m \in \setLines} \sum_{j \in \setLumpyInvestments} \discountFactor\flowMaxDiscrete(\auxBFMuMaxU + \auxBFMuMinU)  \notag\\
&+ \sum_{t \in \setTime} \sum_{n \in \setNodes} \sum_{m \in \setLines} \sum_{i \in \setTariff}  \sum_{k \in \setDemandNode} \discountFactor \tariffDiscrete \PTDF \auxBTauDemand 
\notag\\
&+ \sum_{t \in \setTime} \sum_{n \in \setNodes} \sum_{m \in \setLines} \sum_{i \in \setTariff}  \sum_{p \in \setGenNode} \discountFactor \tariffDiscrete \PTDF\auxBTauGen \geq \notag\\
& \sum_{m \in \setLines} \Big(\uBinary\KFix + \KVar\sum_{j \in \setLumpyInvestments}\bBinaryFlowMax\flowMaxDiscrete\Big)\\
&\eqref{demandQuantityConstraint}-\eqref{flowMinConstraint}\\
&\phiDemMax + \price \geq \priceDemandTilde - \sum_{m \in \setLines} \sum_{i \in \setTariff} \bBinaryTau\tariffDiscrete\PTDF \notag\\
&\hspace{8em} \forall t \in \setTime, \forall n \in \setNodes, \forall k \in \setDemandNode \\
&\phiGenMax - \price \geq - \big(\priceGenTilde + \sum_{m \in \setLines} \sum_{i \in \setTariff} \bBinaryTau\tariffDiscrete\PTDF \big) \notag\\
&\hspace{8em} \forall t \in \setTime, \forall n \in \setNodes, \forall p \in \setGenNode \\
&\eqref{dualFlowConstraint}\\
&\sum_{t \in \setTime} \Big(\sum_{k \in \setDemandAll} \big(\priceDemandTilde\demand - \sum_{m \in \setLines}\sum_{i \in \setTariff}\tariffDiscrete\auxBTauDemand\PTDF\big) \notag\\
&\hspace{3em} - \sum_{p \in \setGenAll} \big(\priceGenTilde\gen + \sum_{m \in \setLines}\sum_{i \in \setTariff}\tariffDiscrete\auxBTauGen\PTDF\big)\Big) = \notag\\
&\sum_{t \in \setTime} \sum_{k \in \setDemandAll} \phiDemMax \demandMax 
+ \sum_{t \in \setTime} \sum_{p \in \setGenAll} \phiGenMax \genMax \notag \\
&+\sum_{t \in \setTime} \sum_{m \in \setLines} \flowZero(\muMax + \muMin) \notag\\
&+ \sum_{t \in \setTime} \sum_{m \in \setLines} \sum_{j \in \setLumpyInvestments} \flowMaxDiscrete(\auxBFMuMaxU + \auxBFMuMinU)\\
&\eqref{conditionsBTau1}-\eqref{conditionsBTau2}\\
&\eqref{auxConstraintBF1}-\eqref{auxConstraintBTauGen2}\label{AppendixMILPLastConstraint}
\end{align}

\bibliographystyle{elsarticle-num}
\bibliography{mybib}

\end{document}